\documentclass[prd,aps,nofootinbib,notitlepage
]{revtex4}
\usepackage{graphicx,epsf,amsfonts,amssymb,amsbsy}
\newcommand{\ds}{\displaystyle}
\newcommand{\vev}[1]{\langle#1\rangle}
\newcommand{\mat}{\left ( \begin{array}}
\newcommand{\emat}{\end{array} \right )}
\newcommand{\vect}{\left ( \begin{array}{c}}
\newcommand{\evect}{\end{array} \right )}

\begin{document}

\title{ 
Dual symmetries of dense three and two-color QCD and some QCD-like NJL models }
\author{T. G. Khunjua $^{1)}$, K. G. Klimenko $^{2)}$, and R. N. Zhokhov $^{3)}$ }
\vspace{1cm}

\affiliation{$^{1}$ The University of Georgia, GE-0171 Tbilisi, Georgia}
\affiliation{$^{2)}$ State Research Center
	of Russian Federation -- Institute for High Energy Physics,
	NRC "Kurchatov Institute", 142281, Protvino, Moscow Region, Russia}
\affiliation{$^{3)}$  Pushkov Institute of Terrestrial Magnetism, Ionosphere and Radiowave Propagation (IZMIRAN),
	108840 Troitsk, Moscow, Russia}

%\pacs{12.39.Ki, 12.38.Mh, 21.65.Qr}
%%% 12.38.Mh Quark-gluon plasma
%%% 21.65.Qr Quark matter
%%% 12.39.Ki Relativistic quark model

%\maketitle
\begin{abstract}
In this paper the symmetry properties of the phase diagram of dense quark matter composed of $u$ and $d$ quarks 
with two or three colors has been investigated in the framework of massless (3+1)-dimensional Nambu--Jona-Lasinio 
(NJL) and QCD models. It turns out that in the presence 
of baryon $\mu_B$, isospin $\mu_I$, chiral $\mu_5$ and chiral isospin $\mu_{I5}$ chemical potentials the Lagrangians 
of these models are invariant under the so-called dual transformations. Consequently, the 
entire NJL model (or QCD) 
thermodynamic potentials are dually symmetric. In particular, it means that in the total 
$(\mu_B,\mu_I,\mu_5,\mu_{I5})$-phase portraits of these models the chiral symmetry breaking (CSB) and charged pion 
condensation (PC) phases are arranged dually conjugated (or symmetrical) to each other (in the case of three-color 
models). Whereas in the case of two-color quark matter, these models predict the entire 
phase structure in which there are
dual symmetries between CSB, charged PC and baryon superfluid phases.
\end{abstract}
%\pacs{12.39.Ki, 12.38.Mh, 21.65.Qr}
%%% 12.38.Mh Quark-gluon plasma
%%% 21.65.Qr Quark matter
%%% 12.39.Ki Relativistic quark model

\maketitle

\section{Introduction}

Earlier, it was established on the basis of some (1+1)- and (2+1)-dimensional massless quantum 
field theory models with
four-fermion interactions (and extended by quark number $\mu$ and chiral $\mu_5$ chemical 
potentials)
that in (toy) dense baryonic matter described in the framework of these low-dimensional 
models
there is a duality correspondence ${\cal D}$ between the chiral symmetry breaking (CSB) and 
superconductivity
phenomena \cite{thies,ebert,cao,ekkz}. As it turned out (see, e.g., in Refs. \cite{ebert,ekkz}),  
this dual correspondence (or symmetry) ${\cal D}$ is based on the invariance of the model 
Lagrangian with respect (i) to the so-called
Pauli-Gursey (PG) transformations of spinor fields $\psi(x)$ \cite{pauli,ojima}, and (ii) simultaneously
performed permutations of the values of the coupling constants in the CSB and superconducting channels, as well as 
values of chemical potentials, $\mu\leftrightarrow\mu_5$. Under the PG transformation the Lagrangian term 
$(\overline{\psi}\psi)^2$ responsible for the
quark-antiquark interaction channel is transformed into the corresponding Lagrangian term $(\psi\psi)^2$ describing
the quark dynamics in the superconducting interaction channel, and vice versa. Moreover, the Lagrangian terms
$\overline{\psi}\gamma^0\psi$ and $\overline{\psi}\gamma^0\gamma^5\psi$ (which are the quark number and chiral densities, respectively),
corresponding to the chemical potentials $\mu$ and $\mu_5$ of the model are also transformed into each other.
Due to this fact, the dual invariance ${\cal D}$ of the Lagrangian leads to the full $(\mu,\mu_5)$-phase diagram of 
dense quark matter, in which CSB and superconducting phases are dually conjugated to each other. It means that they 
are located mirror-symmetrical with respect to the $\mu=\mu_5$ line of the phase diagram. %In Refs. \cite{ebert,ekkz} 
%the dual symmetry ${\cal D}$ of these four-fermion toy models was demonstrated in the framework of 
%the leading order of the large-$N$ expansion technique, where $N$ is the number of spinor fiels. 
In addition, many
characteristics of dense medium, calculated in one of the phases, are easily extrapolated to the dually conjugated
phase of matter. So, the dual symmetry ${\cal D}$ of the model significantly simplifies the investigations of
its thermodynamic properties.

Then, the possibility for duality correspondence between other physical phenomena has been investigated in the 
framework of the massless two-flavor (3+1)-dimensional Nambu--Jona-Lasinio (NJL) model \cite{kkz18} composed of 
$N_c$-colored $u$ and $d$ quarks. This simplest NJL model describes phenomena associated only with quark-antiquark 
interaction channels. In addition, the %The model describes the qIn this case the
Lagrangian of the model was extended by the baryon $\mu_B$, isospin $\mu_I$ and chiral isospin $\mu_{I5}$
chemical potentials, and its thermodynamic potential (TDP) was calculated in the leading order of the large-$N_c$
expansion (or in the mean-field approximation), which describes the phase structure of dense quark matter with
nonzero baryon, isospin and chiral isospin densities. As it was shown in Ref. \cite{kkz18}, in this approximation 
the TDP possesses
the symmetry with respect to the rearrangement of the order parameters of the CSB and the charged pion condensation 
(PC)
phases, as well as the simultaneous transformation of $\mu_I\leftrightarrow\mu_{I5}$. Due to this symmetry of the TDP,
which we call the dual symmetry between CSB and charged PC phases, it turns out that at each fixed
$\mu_B$ the CSB and charged PC phases are arranged mirror-symmetrically relative to the line $\mu_I=\mu_{I5}$ on the
corresponding $(\mu_I,\mu_{I5})$-phase diagram of the model. It means that knowing the phase (and its properties) of
the model at, e.g., $\mu_I<\mu_{I5}$, it is possible to find what a phase (as well as its properties) is realized at
$\mu_I>\mu_{I5}$. Hence, in the mean-field approximation in the above-mentioned simplest NJL model there is a duality 
between
CSB and charged PC phenomena. It is a consequence of the corresponding dual symmetry of the (mean-field) TDP of the 
NJL model. (We emphasize in particular that in the proposed paper, the dual properties of only some (3+1)-dimensional 
models are considered. However, we note that various aspects of the duality between CSB and charged PC phenomena 
were also considered in some (1+1)-dimensional toy models with four-fermion interactions \cite{kkz,Thies2}.)
Then it was shown in Ref. \cite{kkz18-2} that an additional chiral
chemical potential $\mu_5$ does not spoil this duality. However, it must be borne in mind that the duality 
between CSB and 
charged PC is exact (in the large-$N_c$ or mean-field approximations)
only in the chiral limit, i.e. when the bare quark mass $m_0$ is zero. At $m_0\ne 0$ the duality between
CSB and charged PC is only approximative \cite{kkz19}.

In addition, in the recent paper \cite{kkz20} it was shown that in dense baryon matter  with $\mu_B\ne 0$, $\mu_I\ne 0$ and 
$\mu_{I5}\ne 0$, which is formed of two-colored quarks and described by the massless two-flavored and two-colored NJL model, 
two other dual symmetries of its mean-field TDP appear. Recall that in this two-color NJL model an additional diquark interaction 
channel presents (see, e.g., in Refs. \cite{kogut,son2,weise,ramos,andersen3,brauner1,andersen2,imai,chao,Astrakhantsev:2020tdl}).
As a result, the dualities between CSB and diquark condensation as well as between charged PC and diquark condensation phenomena appear, 
in addition to the above-mentioned duality between CSB and charged PC. In this NJL model, which is a low-energy effective model of 
the two-color QCD with two flavors, the investigation of dense quark matter was also 
performed only in the mean-field approximation.

And quite recently the dual symmetries of a more involved three-color and two-flavor NJL model with additional 
diquark interaction channel have been investigated in the chiral limit and also using the mean-field approximation 
\cite{u2}. It was shown in this work that, despite the presence of a phase with color superconductivity, 
%despite the fact %that in the presence of several chemical potentials 
%that the model describes dense quark matter with an additional color superconducting phase, 
in the phase portrait of the model there is a dual symmetry only between the CSB and the charged PC phases 
(the phase with color superconductivity transforms into itself under this dual transformation). 
Hence, we see that two different three-color NJL models that effectively describe (in the general case, different) 
low-energy QCD regions have thermodynamic potentials that are invariant in the chiral limit and in the mean-field 
approximation under the same dual transformation relating the phenomena CSB and charged PC.

Now, two questions naturally arise. (i) Whether the dual relations between various physical phenomena are inherent in 
the above-mentioned 
dense NJL models as a whole, and not only in their mean-field approximations. In addition, (ii) one can pose the 
question even more 
broadly and try to clarify the situation with the dual symmetries within the dense QCD themselves,
both with two and three colors of quarks. The present paper is devoted to the consideration of these questions.

We show that massless QCD and QCD-like effective NJL Lagrangians, describing dense baryonic matter composed of two- and
three-colored quarks, are invariant with respect to some duality transformations. Due to this fact, the confidence
appears that the corresponding full TDPs are dually symmetric. It means that dualities between different phenomena
of dense baryonic medium, which were previously
observed in the mean-field approximation and in the leading large-$N_c$ order, should be realized also outside the
mean-field approximation (or in other orders of the $1/N_c$ expansion). Thus, the dualities observed in
Refs. \cite{kkz18,kkz18-2,kkz,kkz19,kkz20} are not specific only
to the approximation method of research (including the order of approximation), as some might think, but
they are inherent in the QCD or the corresponding NJL model itself, on the basis of which dense quark matter is studied.

The paper is organized as follows. In section \ref{a}, we introduce the notion of the dual transformation and what 
is the difference between it and the usual symmetry transformation in field theory. Then it is shown in this section 
that in the simplest two-flavor NJL model with massless three-colored quarks, which describes dense quark matter 
with isospin and chiral imbalances, there is a duality between CSB  and charged PC phenomena. It means
that the Lagrangian of the model is symmetric under the discrete ${\cal D}_1$ dual transformation, when the 
four-fermion structures of the CSB and charged PC interaction channels are transformed to each other by some 
Pauli-Gursey-type transformation of quark fields, as well as after the transmutation
$\mu_I\leftrightarrow\mu_{I5}$. In addition, the order parameters characterizing each of these phases are also 
conjugated to  %dually ${\cal D}_1$ conjugated to
each other under the action of ${\cal D}_1$. Then in section \ref{b} we show that the Lagrangian of a more extended 
NJL model with additional diquark interaction channel is also invariant under the same dual transformation 
${\cal D}_1$. In section III the duality properties of the two-color massless NJL model is investigated when it is 
also extended by the $\mu_B$, $\mu_I$, $\mu_{I5}$ and $\mu_5$ chemical potential terms. We show
that in addition to the dual ${\cal D}_1$, this NJL model is invariant under two new dual transformations. 
One of them, ${\cal D}_2$,  includes the transmutation $\mu_B\leftrightarrow\mu_I$ and the PG-type mapping of quark 
fields that connects the four-quark interaction structures of the charged PC and diquark channels, i.e. there appears 
a duality correspondence between charged PC and baryon superfluid (BSF) phenomena. Another one,
${\cal D}_3$, is constructed from the transmutation $\mu_B\leftrightarrow\mu_{I5}$ and PG-type transformation of 
quarks that connects the four-quark structures in the CSB and diquark channels. It means that in the two-color NJL 
model there is a duality between CSB and BSF. In section IV it is shown that the two-flavor QCD Lagrangians 
composed of massless both three- and two-color quarks and extended by the above mentioned chemical potential terms 
are symmetric with respect to the ${\cal D}_1$ (in the three-color case) and ${\cal D}_1$, ${\cal D}_2$ and 
${\cal D}_3$ dual transformations (in the two-color case).

\section{Dual symmetry between CSB and charged PC in some three-color NJL models}
\subsection{The case of the simplest NJL model} \label{a}

In the recent paper \cite{kkz18-2} the phase structure of dense quark matter in thermodynamic equilibrium was 
investigated at zero temperature ($T=0$) in the mean-field approximation (or in the leading order of the $1/N_c$ 
expansion) in the framework of the three-color and two-flavor massless NJL model with four-fermion Lagrangian of 
the following form
\begin{equation}
L_{NJL}=i\overline{\psi} \gamma^{\mu}\partial_{\mu} \psi+\overline{\psi}\gamma^0{\cal M}\psi+
G\Big\{(\overline{\psi}\psi)^2+ (i\overline{\psi}\gamma^5\vec\tau\psi)^2\Big\},\label{1}
\end{equation}
where 
\begin{eqnarray}
{\cal M}=\frac{\mu_B}{3}+\frac{\mu_I}2\tau_3+\frac{\mu_{I5}}2\gamma^5\tau_3+\mu_5\gamma^5
  \label{200}
\end{eqnarray}  
and $\psi\equiv\psi(x)$ is a flavor doublet, i.e. $\psi=\left( {\begin{array}{c}
\psi_{u} \\
\psi_d
\end{array} } \right)$. In addition, it is a color triplet (or, alternatively, color doublet as in the
sections below) as well; $\tau_k$ ($k=1,2,3$) are the Pauli matrices acting in two-dimensional flavor space. The Lagrangian (1) contains baryon $\mu_B$, isospin $\mu_I$, chiral isospin $\mu_{I5}$,
and chiral $\mu_{5}$ chemical potentials. So it is intended to describe the properties of quark matter (throughout the paper we consider only its state of thermodynamic equilibrium) with nonzero
baryon $n_B=(n_u+n_d)/3$ and isospin $n_I=(n_u-n_d)/2$ densities (here $n_u$ and $n_d$ are densities of the $u$ and $d$ quarks, respectively). These quantities are thermodynamically conjugated to baryon $\mu_B$ and isospin $\mu_I$ chemical potentials. Moreover, it is supposed that in dense quark matter there is a nonzero difference between the densities $n_R$ and $n_L$ of all right- and all left-handed quarks. Hence, the chiral density of the system, i.e.
the quantity $n_5\equiv n_R-n_L$ and its thermodynamical conjugation, the chiral chemical potential
$\mu_5$, are nonzero. Finally, we suppose also that chiral densities $n_{u5}$ and $n_{d5}$ of $u$ and $d$ quarks,
respectively, are different. It means that chiral isospin density $n_{I5}\equiv n_{u5}-n_{d5}$ of the system is
nonzero (note that $n_{5}= n_{u5}+n_{d5}$), which is thermodynamically conjugated to the chiral isospin chemical 
potential $\mu_{I5}$ of quark matter.

The quantities $n_B$, $n_I$ and $n_{I5}$ are densities of conserved charges, which correspond to the invariance of
Lagrangian (\ref{1}) with respect to the abelian $U_B(1)$, $U_{I_3}(1)$ and $U_{AI_3}(1)$
groups, where \footnote{\label{f1,1}
Recall for the following that~~
$\exp (\mathrm{i}\alpha\tau_3)=\cos\alpha
+\mathrm{i}\tau_3\sin\alpha$,~~~~
$\exp (\mathrm{i}\alpha\gamma^5\tau_3)=\cos\alpha
+\mathrm{i}\gamma^5\tau_3\sin\alpha$.}
\begin{eqnarray}
U_B(1):~\psi\to\exp (\mathrm{i}\alpha/3)\psi;~
U_{I_3}(1):~\psi\to\exp (\mathrm{i}\alpha\tau_3/2)\psi;~
U_{AI_3}(1):~\psi\to\exp (\mathrm{i}
\alpha\gamma^5\tau_3/2)\psi.
\label{2}
\end{eqnarray}
We emphasize that the Lagrangian (\ref{1}) is invariant under each of the transformations in Eq. (\ref{2})
and that, in this case, the external parameters of the system, i.e. chemical potentials, coupling constants, etc., remain unchanged.
So we have from (\ref{2}) that $n_B=\vev{\overline{\psi} \gamma^0\psi}/3$, $n_I=\vev{\overline{\psi}\gamma^0\tau^3
\psi}/2$ and $n_{I5}=\vev{\overline{\psi}\gamma^0\gamma^5\tau^3\psi}/2$, where $\vev{\cdots}$ means the ground state
(or the state of thermodynamic equilibrium) expectation value. However, the chiral chemical potential $\mu_5$ does not correspond to a conserved
quantity of the model (\ref{1}). It is usually introduced in order to describe a system on the time scales,
when all chirality changing processes are finished in the system, so it is in the state of thermodynamical
equilibrium with some fixed value of the chiral density $n_5$ \cite{andrianov}.

Finally, before proceeding to a more detailed consideration of the NJL model (\ref{1}), we should note that cold 
dense quark matter with isospin asymmetry can exist in the cores of neutron stars. Usually, see, e.g., in Refs. 
\cite{son,he,ak,ekkz2,Andersen:2018nzq,Ayala}, 
such quark medium is studied using effective models involving only two chemical potentials, $\mu_B$ and $\mu_I$.
But this may not be enough to describe its real properties. The fact is that neutron stars have very strong magnetic 
fields, due to which a nonzero chiral $n_5$ and chiral isospin $n_{I5}$ densities can be generated in dense quark 
medium on the basis of the chiral magnetic \cite{fukus} and chiral separation \cite{Metlitski} effects 
(see the corresponding discussion in Refs. \cite{kkz19,Khun}). Therefore, we think that within the framework of model (\ref{1}) with four chemical potentials, 
one can obtain a more adequate description of the properties of dense quark matter.

To simplify the investigation of the phase structure of the NJL model (\ref{1}), usually the auxiliary $\widetilde L_{NJL}$ Lagrangian is used,
\begin{eqnarray}
\widetilde L_{NJL}\ds &=&\overline \psi\Big [\gamma^\rho\mathrm{i}\partial_\rho
+\mu\gamma^0+\mu_5\gamma^0\gamma^5
+ \nu\tau_3\gamma^0+\nu_{I5}\tau_3\gamma^0\gamma^5-\sigma
-\mathrm{i}\gamma^5\pi_a\tau_a\Big ]\psi
 -\frac{1}{4G}\Big [\sigma^2+\vec\pi^2\Big ],
\label{3}
\end{eqnarray}
where $\mu=\mu_B/3$, $\nu=\mu_I/2$, $\nu_5=\mu_{I5}/2$.

Starting from the auxiliary Lagrangian (\ref{3}), it is possible in principle to obtain %a functional $\Gamma (\sigma, \vec\pi)$ which is 
a generating functional $\Gamma (\sigma, \vec\pi)$ of the one-particle-irreducible Green functions of the $\sigma (x)$ and $\vec\pi (x)$
fields (see, e.g., in \cite{iliopoulos}). Usually, $\Gamma (\sigma, \vec\pi)$ is also called an effective action of the model and allows us
to introduce the quantity $\Omega (\sigma,\vec\pi)$ which is called the full thermodynamic potential (TDP) of the model
(but at zero chemical potentials, it is usually called the full effective potential of the model),
\begin{eqnarray}
\Omega(\sigma,\vec\pi)\int d^4x\equiv -\Gamma (\sigma,\vec\pi)\Big |_{\sigma,\pi_1,\pi_2,\pi_3 =const}
\label{7}
\end{eqnarray}
The coordinates of the global minimum point (GMP) of the TDP $\Omega(\sigma,\vec\pi)$ are the ground state
expectation values $\vev{\sigma}$ and $\vev{\pi_a}$ of the auxiliary bosonic fields, which are no more
than the order parameters whose behaviors vs chemical potentials define the phase structure of the model. Hence, if,
e.g., in the GMP we have  $\vev{\sigma}\ne 0$ and/or $\vev{\pi_3}\ne 0$ but $\vev{\pi_{1,2}}=0$, then, as it is
clear from Eq. (\ref{5}), the $U_{AI_3}(1)$ chiral symmetry is spontaneously broken down and the CSB phase is
realized in quark matter. If $\vev{\sigma}= 0$, $\vev{\pi_3}=0$ but $\vev{\pi_{1,2}}\ne 0$, then isospin $U_{I_3}(1)$
invariance of the model is spontaneously broken and the
charged PC phase is observed there, etc. So the investigation of the TDP  $\Omega(\sigma,\vec\pi)$ on the least value vs $\sigma$ and $\vec\pi$ is a very important task. But in order to simplify the solution of the problem, note that it is clear that $\sigma$ and $\vec\pi$ are
transformed under the isospin $U_{I_3}(1)$ and axial isospin (or chiral) $U_{AI_3}(1)$ groups in the following manner:
\begin{eqnarray}
U_{I_3}(1):~&&\sigma\to\sigma;~~\pi_3\to\pi_3;~~\pi_1\to\cos
(\alpha)\pi_1+\sin (\alpha)\pi_2;~~\pi_2\to\cos (\alpha)\pi_2-\sin
(\alpha)\pi_1,\nonumber\\
U_{AI_3}(1):~&&\pi_1\to\pi_1;~~\pi_2\to\pi_2;~~\sigma\to\cos
(\alpha)\sigma+\sin (\alpha)\pi_3;~~\pi_3\to\cos
(\alpha)\pi_3-\sin (\alpha)\sigma.
\label{5}
\end{eqnarray}
As a consequence, for each fixed set $\mu,\mu_I,\mu_{I5}$ and $\mu_5$ of chemical potentials the TDP
$\Omega(\sigma,\vec\pi)$ is really a function of only two field combinations, $\Sigma\equiv\sqrt{\sigma^2+\pi_3^2}$ and $\Pi\equiv\sqrt{\pi_1^2+\pi_2^2}$, i.e.
\begin{eqnarray}
\Omega(\sigma,\vec\pi)\equiv \Omega \big(\Sigma,\Pi\big).
\label{8}
\end{eqnarray}
In this case to find the phase structure of the model, it is enough to investigate (for each fixed set of chemical potentials)
the function $\Omega (\Sigma,\Pi)$ on the global minimum point vs. $\Sigma$ and $\Pi$.
%The coordinates of the GMP are the order parameters, and just their behavior vs. chemical potentials define the $(\mu,\mu_I,\mu_{I5},\mu_5)$-phase portrait of the model (or the phase structure of dense quark matter with isospin and chiral asymmetry).
So, if in the GMP of the function (\ref{8}) vs $\Sigma$ and $\Pi$
we have $(\Sigma\ne 0,\Pi=0)$, then the phase with spontaneous breaking of the chiral symmetry $U_{AI_3}(1)$ is realized in
the system. If in the GMP of the TDP (\ref{8}) we have $(\Sigma =0, \Pi\ne 0)$, then charged PC phase is observed
(in this case the isospin $U_{I_3}(1)$ symmetry is spontaneously broken) and,
finally, the GMP of the form $(\Sigma=0,\Pi=0)$ corresponds to the symmetrical phase of the model, in which none of the symmetry groups (\ref{2}) is broken spontaneously. \footnote{Indeed, if, e.g.,  in the GMP of the TDP $\Omega (\Sigma,\Pi)$ we have $(\Sigma\ne 0,\Pi=0)$,
it means that in this case there exist nonzero values $\sigma=\vev{\sigma}$ and/or $\pi_3=\vev{\pi_3}$ corresponding to the least value of the TDP vs $\sigma$ and $\vec\pi$, i.e. of the function $\Omega(\sigma,\vec\pi)$.  Since the quantities $\vev{\sigma}$ and/or $\vev{\pi_3}$ are not invariant under the transformations from $U_{AI_3}(1)$ group (as it follows from Eq. (\ref{5})), the chiral symmetry is spontaneously broken down in this case, etc. \label{f3}}

In the recent paper \cite{kkz18-2} the TDP (\ref{8}) of the model has been investigated in the mean-field
(or in the leading large-$N_c$ order) approximation (in this case we use for it the notation
$\Omega_{mf}(\Sigma,\Pi)$). Moreover, it was shown there that $\Omega_{mf}(\Sigma,\Pi)$ is a quantity which is invariant under the transformation
\begin{eqnarray}
\widetilde{\cal D}_1:~~\Sigma \longleftrightarrow\Pi,~~~\mu_{I} \longleftrightarrow \mu_{I5}.\label{9}
\end{eqnarray}
Due to this property of the TDP in the mean-field approximation, %$\Omega_{mf}\big(\Sigma,\Pi\big)$,
it was shown in Ref. \cite{kkz18-2} that
there is a duality between CSB and charged PC phenomena (so we call the $\widetilde{\cal D}_1$ (\ref{9}) as a dual transformation).
It means that at fixed $\mu$ and $\mu_5$ the $(\mu_I,\mu_{I5})$-phase portrait (obtained in the framework of the 
mean-field approximation) of the NJL model (\ref{1}) obeys a symmetry with
respect to simultaneous transformations, CSB$\leftrightarrow$charged PC and $\mu_{I}\leftrightarrow\mu_{I5}$.
In more detail, suppose that at some fixed particular values of the chemical potentials $\mu$, $\mu_{5}$, $\mu_I=A$ and $\mu_{I5}=B$
the global minimum of the TDP $\Omega_{mf}(\Sigma,\Pi)$ lies at the point, e.g.,
$(\Sigma=\Sigma_0\ne 0,\Pi=0)$. In this case, for such fixed values of the chemical potentials the CSB phase is
realized in the model. Then it follows from the invariance of the TDP with respect to the duality transformation
$\widetilde{\cal D}_1$ (\ref{9}) that at permuted chemical potential values (i.e. at $\mu_I=B$ and $\mu_{I5}=A$ and
intact values of $\mu$ and $\mu_{5}$) the global minimum of the TDP $\Omega_{mf}(\Sigma,\Pi)$ is arranged
at the point $(\Sigma=0,\Pi=\Sigma_0)$, which corresponds to the charged PC phase (and vice versa). 
Hence, the knowledge of a phase of the model (\ref{1}) at some fixed values of external free model parameters
$\mu,\mu_I,\mu_{I5},\mu_5$ is sufficient to understand what a phase (we call it a dually conjugated) is realized at
rearranged values of external parameters, $\mu_I\leftrightarrow\mu_{I5}$, at fixed $\mu$ and $\mu_{5}$. Moreover,
different physical parameters such as condensates, densities, etc, which characterize both the initial phase and
the dually conjugated one, are connected by the duality transformation $\widetilde{\cal D}_1$. For example, the chiral
condensate of the initial CSB phase at some fixed $\mu,\mu_I,\mu_{I5},\mu_5$ is equal to the charged-pion condensate
of the dually conjugated charged PC phase, in which one should perform the replacement $\mu_I\leftrightarrow\mu_{I5}$.
Knowing the particle density $n_q(\mu_I,\mu_{I5})$ of the initial CSB phase as a function of chemical potentials
$\mu_I,\mu_{I5}$, one can find the particle density in the dually conjugated charged PC phase by interchanging
$\mu_I$ and $\mu_{I5}$ in the expression $n_q(\mu_I,\mu_{I5})$, etc.

Thus, the symmetry of the TDP of the massless NJL model (\ref{1}) with respect to the dual transformation (\ref{9})
makes it possible to significantly simplify the study of the properties of dense ($\mu_B\ne 0$) quark matter with
isospin ($\mu_I\ne 0$) and chiral isospin ($\mu_{I5}\ne 0$) asymmetries. However, it is worth emphasizing once again that the
resulting duality between CSB and charged PC phenomena was established in Ref. \cite{kkz18-2} only 
on the basis of the TDP calculated within the framework of
the mean-field (or in the leading order of the large-$N_c$) approximation. Is this duality property inherent in
the NJL model as a whole, i.e. outside the mean-field (or in other orders of the $1/N_c$) approximation, remains
formally unclear. 

In the present paper, we show that (i) there is a discrete Pauli--Gursey-type transformation of
quark fields \footnote{The most general definition of the Pauli--Gursey transformation is given below Eq. 
(\ref{17}). } that transforms the Lagrangian (\ref{1}) terms responsible for the CSB interaction channel into 
a term of an interaction channel in which condensation of charged pions occurs (and vice versa). In addition, (ii) in this case, the Lagrangian (\ref{1}) terms which are the density operator of the isospin charge and the density operator of the chiral isospin charge are transformed to each other (other terms of the NJL Lagrangian (\ref{1}) remain intact).
Hence, (iii) if $\mu_I\leftrightarrow\mu_{I5}$ simultaneously with this Pauli--Gursey-type transformation, the Lagrangian
(\ref{1}) as a whole remains unchanged. As a consequence, it turns out that the dual symmetry (\ref{9}) is inherent 
to the full thermodynamic potential $\Omega (\Sigma,\Pi)$ of the system, and not only to its mean-field approximation
$\Omega_{mf} (\Sigma,\Pi)$.

Let us define the following transformation of the flavor doublet $\psi(x)$ (as it is clear from the following consideration, it is indeed a Pauli-Gursey transformation),
\begin{eqnarray}
PG_1:~~\psi_R\to\psi_R,~~~\psi_L\to i\tau_1\psi_L,~~ {\rm or}~~\overline{\psi_R}\to \overline{\psi_R},~~
\overline{\psi_L}\to \overline{\psi_L} (-i)\tau_1,\label{10}
\end{eqnarray}
where $\psi=\psi_L+\psi_R$ and
\begin{eqnarray*}
\psi_R\equiv\frac{1+\gamma^5}2\psi\equiv\Pi_+\psi,~~~\psi_L\equiv\frac{1-\gamma^5}2\psi\equiv\Pi_-\psi,~~{\rm i.e.}
~~\overline{\psi_R}=\overline\psi\Pi_-,~~\overline{\psi_L}=\overline\psi\Pi_+.\label{11}
\end{eqnarray*}
Alternatively, it can be presented in the following form
\begin{eqnarray}
PG_1:~~\psi\to \psi_R+i\tau_1\psi_L,~~~{\rm i.e.}~~~ PG_1\psi\equiv PG_1\left( {\begin{array}{c}
	 \psi_{u} \\
	 \psi_d
	\end{array} } \right)=\left( {\begin{array}{c}
	 \psi_{uR}+i\psi_{dL} \\
	 \psi_{dR}+i\psi_{uL}
	\end{array} } \right).\label{12}
\end{eqnarray}
Using Eqs. (\ref{10})-(\ref{12}), it is easy to find out how the simplest quark-antiquark structures of the Lagrangian (\ref{1}) are transformed under the action of the Pauli--Gursey-type transformation $PG_1$,
\begin{eqnarray}
\overline{\psi}\psi=\overline{\psi_{R}}\psi_{L}+\overline{\psi_{L}}\psi_{R}&\stackrel{PG_1}{\longrightarrow}&
\overline{\psi_{R}}i\tau_{1}\psi_{L}-
\overline{\psi_{L}}i\tau_{1}\psi_{R}=-i \overline{\psi}\gamma^{5}\tau_{1}\psi\label{13}\\
%\end{equation}
%\begin{equation}
i\overline{\psi}\gamma^{5}\tau_{1}\psi=i(-\overline{\psi_{R}}\tau_{1}\psi_{L}+
\overline{\psi_{L}}\tau_{1}\psi_{R})
&\stackrel{PG_1}{\longrightarrow}&
-i\overline{\psi_{R}}\tau_{1}i\tau_{1}\psi_{L}-i\overline{\psi_{L}}i\tau_{1}\tau_{1}\psi_{R}=\overline{\psi}\psi.
\label{14}
\end{eqnarray}
In a similar way it is possible to show that
\begin{eqnarray}
i \overline{\psi}\gamma^{5}\tau_{2}\psi%=i(-\bar{\psi_{L}}\tau_{2}\psi_{R}+\bar{\psi_{R}}\tau_{2}\psi_{L})
&\stackrel{PG_1}{\longrightarrow}&%-i\bar{\psi_{L}}\tau_{2}i\tau_{1}\psi_{R}-
%i\bar{\psi_{R}}i\tau_{1}\tau_{2}\psi_{L}=-i\bar{\psi_{L}}\tau_{3}\psi_{R}+i\bar{\psi_{R}}\tau_{3}\psi_{L}=
i\overline{\psi}\gamma^{5}\tau_{3}\psi,\label{15}\\
%\end{equation}
%\begin{equation}
i \overline{\psi}\gamma^{5}\tau_{3}\psi %=i(-\bar{\psi_{L}}\tau_{3}\psi_{R}+\bar{\psi_{R}}\tau_{3}\psi_{L})
&\stackrel{PG_1}{\longrightarrow}&
%-i\bar{\psi_{L}}\tau_{3}i\tau_{1}\psi_{R}-i\bar{\psi_{R}}i\tau_{1}\tau_{3}\psi_{L}=
%i\bar{\psi_{L}}\tau_{2}\psi_{R}-i\bar{\psi_{R}}\tau_{2}\psi_{L}=
-i\overline{\psi}\gamma^{5}\tau_{2}\psi.\label{16}
\end{eqnarray}
It means that four-fermion structures of the Lagrangian (\ref{1}) that are responsible for the CSB and charged PC
are transformed to each other by the $PG_1$ transformation (\ref{10})-(\ref{12}), i.e.
\begin{equation}
(\overline{\psi}\psi)^2+
(i\overline{\psi}\gamma^5\tau_3\psi)^2\stackrel{PG_1}{\longleftrightarrow} (i\overline{\psi}\gamma^5\tau_1\psi)^2
+(i\overline{\psi}\gamma^5\tau_2\psi)^2. \label{17}
\end{equation}
(Note that the original Pauli--Gursey transformation of Ref. \cite{pauli} connects the four-fermion structures of 
low-dimensional Lagrangians responsible for the CSB and superconducting channels (see, e.g., in Refs. 
\cite{ebert,cao,ekkz,ojima}). So, in order to highlight this particular feature of the PG transformation, in the present paper {\bf we call any transformation of spinor fields that connects the quark structures of various interaction channels to each other as a Pauli--Gursey-type one.})
Since the free term $i\overline{\psi} \gamma^{\mu}\partial_{\mu} \psi$ of the Lagrangian (\ref{1}) remains intact under the action of the $PG_1$ transformation,
it is clear that at zero chemical potentials the initial NJL Lagrangian (\ref{1}) is invariant under the Pauli--Gursey-type
transformation (\ref{10})-(\ref{12}). However, at nonzero chemical potentials the $PG_1$ is no more the symmetry
transformation of the NJL model. Indeed, under the action of $PG_1$ the quantities
$\overline{\psi}\gamma^{0}\psi$ and $\overline{\psi}\gamma^{0}\gamma^{5}\psi$ are not changed, but
\begin{eqnarray}
\overline{\psi}\gamma^{0}\gamma^{5}\tau_{3}\psi&=&(\overline{\psi_{R}}\gamma^{0}\tau_{3}\psi_{R}-
\overline{\psi_{L}}\gamma^{0}\tau_{3}\psi_{L})\stackrel{PG_1}{\longleftrightarrow}(\overline{\psi_{R}}\gamma^{0}\tau_{3}\psi_{R}+
\overline{\psi_{L}}i\tau_{1}\gamma^{0}\tau_{3}i\tau_{1}\psi_{L})\nonumber\\&=&(\overline{\psi_{R}}\gamma^{0}\tau_{3}\psi_{R}+
\overline{\psi_{L}}\gamma^{0}\tau_{3}\psi_{L})=\overline{\psi}\gamma^{0}\tau_{3}\psi.\label{18}
\end{eqnarray}
It means that densities of isospin and chiral isospin charges, i.e. the quantities $\overline{\psi}\gamma^{0}\tau_{3}\psi$ and $\overline{\psi}\gamma^{0}\gamma^{5}\tau_{3}\psi$ of Eq. (\ref{18}), go one into another and, as a result, at $\mu_I\ne 0$ and $\mu_{I5}\ne 0$ the NJL Lagrangian (\ref{1}) is not invariant under the $PG_1$
transformation (\ref{10})-(\ref{12}). Hence, at nonzero $\mu_I\ne 0$ and $\mu_{I5}\ne 0$ the discrete $PG_1$ transformation
is no more the symmetry transformation of the model. But it is evident that if simultaneously with field transformation
(\ref{10})-(\ref{12}) we perform the permutation of the chemical potentials, $\mu_I\longleftrightarrow\mu_{I5}$, then
the NJL Lagrangian remains invariant. {\bf Transformations under which not only the field variables of the Lagrangian, but
simultaneously with them also some of the external parameters (to which we include chemical potentials, coupling
constants, etc.)  are changed, we will further call dual transformations}. Thus, the NJL model (\ref{1}) is invariant
(or symmetric) with respect to the dual ${\cal D}_1$ transformation, where
\begin{eqnarray}
{\cal D}_1:~~\psi\to \psi_R+i\tau_1\psi_L,~~~\mu_I\longleftrightarrow\mu_{I5}.\label{20}
\end{eqnarray}
Since in this case the quark structures of CSB and charged PC channels of interaction pass into each other (see, e.g., in Eq.
(\ref{17})), we will say that each of these channels is ${\cal D}_1$ dual to the other, or that the CSB and charged PC phenomena
are dually ${\cal D}_1$ conjugated to each other in the framework of the NJL model (\ref{1}).

%Recall that the full TDP (\ref{7})-(\ref{8}) of the NJL model is obtained on the basis of the auxiliary Lagrangian (\ref{3}). So let us consider how these quantities are transformed by the action of ${\cal D}_1$. % And we are going to show that the auxiliary NJL Lagrangian (\ref{3}) is also ${\cal D}_1$ invariant.
It is clear that under the $PG_1$ transformation (\ref{10})-(\ref{12}) (and also under the action 
of ${\cal D}_1$ (\ref{20}))  $\sigma$ and $\vec\pi_a$ are transformed 
by $PG_1$ (or by ${\cal D}_1$) in the following way
\begin{eqnarray}
PG_1,~{\cal D}_1:~~\sigma\to-\pi_1,~~\pi_1\to\sigma,~~\pi_2\to\pi_3,~~\pi_3\to-\pi_2.
\label{21}
\end{eqnarray}
Hence, in this case the quantities $\Sigma$ and $\Pi$ (see the definition before Eq. (\ref{8})) are transformed into each other, 
$\Sigma\leftrightarrow\Pi$. %Moreover, it follows from Eq. (\ref{21}) that the term of the auxiliary NJL Lagrangian (\ref{3}), which is proportional to $(\sigma^2+\vec\pi^2)$, is $PG_1$ and ${\cal D}_1$ invariant. In addition, taking into account the transformation rules (\ref{10}) and (\ref{21}), one can see that the Yukawa-like term $\overline \psi (\sigma +\mathrm{i}\gamma^5\pi_a\tau_a )\psi$ of this Lagrangian, as well as its free term $i\overline{\psi} \gamma^{\mu}\partial_{\mu} \psi$, also remain intact under the action of $PG_1$. So at zero chemical potentials the Lagrangian $\widetilde L_{NJL}$ (\ref{3}) is $PG_1$ symmetric (however, it is not the case when $\mu_I$ and $\mu_{I5}$ are nonzero). Finally, taking into account the relations (\ref{18}) and (\ref{21}), it is easy to see that at nonzero chemical potentials, auxiliary Lagrangian (\ref{3}) as a whole is symmetric with respect to the dual ${\cal D}_1$ transformation (\ref{20}), in which it is necessary to add the ${\cal D}_1$ transformation of the scalar and pseudoscalar fields (\ref{21}). 
The 
full effective 
action $\Gamma(\sigma,\vec\pi)$ and the full TDP $\Omega (\Sigma,\Pi)$ (\ref{8}) of the 
model will be invariant with respect to the same discrete dual 
transformation ${\cal D}_1$ (\ref{20})-(\ref{21}).
But the last property, i.e. the ${\cal D}_1$ symmetry of the TDP, 
is nothing else than its invariance under the dual transformation $\widetilde{\cal D}_1$ 
(\ref{9}), when 
$\Sigma\leftrightarrow\Pi$ and $\mu_I\leftrightarrow\mu_{I5}$ (see also the remark just below 
Eq. (\ref{21})). As a result, the dual symmetry between CSB and charged PC phenomena of dense quark medium with 
isospin and chiral 
isospin asymmetries, %(see the description below Eq. (\ref{9})), 
predicted earlier in Ref. \cite{kkz18-2} within the 
framework of the mean-field  approximation to the phase structure, is in fact a genuine property of dense quark matter described by 
the massless NJL model (\ref{1}). So the duality between CSB and charged PC phenomena is realized in the full 
$(\mu,\mu_I,\mu_{I5},\mu_5)$-phase diagram of this model (and not only in the scope of the mean-field approximation), 
since it is a consequence of the dual symmetry ${\cal D}_1$ (\ref{20}) of the microscopic NJL Lagrangian (\ref{1}).

\subsection{Dual symmetry of the NJL model extended by diquark interaction channel }\label{b}

Recall that in the recent paper \cite{u2} the phase structure of the four-fermion model, 
whose Lagrangian $L_{CSC}$,
\begin{equation}
L_{CSC}=L_{NJL}+H\sum_{A=2,5,7}
[\overline{\psi^c}i\gamma^5\tau_2\lambda_{A}\psi]
[\overline{\psi} i\gamma^5\tau_2\lambda_{A} \psi^c],\label{100}
\end{equation}
differs from Eq. (\ref{1}) by an additional term with diquark interaction, was discussed. 
In Eq. (\ref{100}),  $\lambda_{A}$ $(A=2,5,7)$ are the Gell-Mann 3$\times$3 matrices acting in the 
three-dimensional fundamental representation of the color $SU(3)_c$ group; $\psi^c=C\overline\psi^T$, 
$\overline{\psi^c}=\psi^T C$ are charge-conjugated spinors, where $C=i\gamma^2\gamma^0$ is
the charge conjugation matrix (the symbol $T$ denotes the transposition operation). Other notations are the same 
as for the Lagrangian $L_{NJL}$ (see the text below Eq. (\ref{1})). As in the case with the simpler NJL model from 
the previous subsection, when considering the phase structure of the model (\ref{100}) it is
also convenient to deal with its auxiliary Lagrangian $\widetilde L_{CSC}$, 
\begin{eqnarray}
\widetilde L_{CSC}%\ds 
&=&\widetilde L_{NJL}-\frac1{4H}\Delta^{*}_{A}\Delta_{A}-
 \frac{\Delta^{*}_{A}}{2}[\overline{\psi^c}i\gamma^5\tau_2\lambda_{A} \psi]
-\frac{\Delta_{A'}}{2}[\overline\psi i\gamma^5\tau_2\lambda_{A'}\psi^c],
\label{CSC}
\end{eqnarray}
where $\widetilde L_{NJL}$ is an auxiliary semibosonized Lagrangian (\ref{3}). In addition to spinor fields $\psi(x)$
and auxiliary bosonic fields $\sigma(x),\vec\pi(x)$, 
$\widetilde L_{CSC}$ contains three auxiliary bosonic 
diquark fields $\Delta_A(x)$, where $A=2,5,7$.
Their ground state
expectation values are the order parameters of the color superconducting phase. \footnote{The details of the construction of the auxiliary 
semi-bosonic Lagrangian $\widetilde L_{CSC}$ 
are described  both in the paper \cite{u2} and in earlier reviews devoted to color superconductivity \cite{buballa}.} 
On the basis of the microscopic auxiliary Lagrangian (\ref{CSC}), it is much easier to construct formally the full TDP
$\Omega(\sigma,\pi_a,\Delta_A)$ of the model (\ref{100}), which is indeed a function of constant 
scalar fields $\sigma,\pi_i,\Delta_A$. So if in the GMP of this total TDP at least one of the quantities $\Delta_A$ is not equal to zero, then color $SU(3)_c$ symmetry 
of the model (\ref{100}) is spontaneously broken, the diquark pairing occurs,  and the so-called color 
superconducting phase is realized. Of course, due to the Abelian $U_{I_3}(1)$ and $U_{AI_3}(1)$ symmetries (\ref{2}) 
of Lagrangian (\ref{100}), as well as its invariance with respect to the $SU(3)_c$ group, the total TDP of this
model depends only on the quantities $\Sigma$, $\Pi$ (these quantities are defined around Eq. (\ref{8})), as well 
as on $\Delta\equiv\sqrt{\Delta_2\Delta^*_2+\Delta_5\Delta^*_5+\Delta_7\Delta^*_7}$, i.e. 
$\Omega(\sigma,\pi_a,\Delta_A)\equiv\Omega(\Sigma,\Pi,\Delta)$. 

Finally, it is necessary to note that in Ref. \cite{u2} not the full TDP $\Omega(\Sigma,\Pi,\Delta)$ of the model 
(\ref{100})-(\ref{CSC}), but only its mean-field approximation $\Omega_{mf}(\Sigma,\Pi,\Delta)$ has been investigated.
And within the framework of this approach, the invariance of $\Omega_{mf}(\Sigma,\Pi,\Delta)$ with respect to the 
transformation $\widetilde{\cal D}_1$ (\ref{9}) was discovered (it is supposed in this case that $\Delta$ 
is unchanged). 
Hence, it was established in the mean-field approximation that in quark matter at rather high baryon
densities, i.e. when CSC should be taken into account, there is a duality between CSB and charged PC phenomena. In the present 
paper, we show that this kind of dual symmetry is an inherent feature of quark matter, whose dynamics is 
described by Lagrangian (\ref{100})-(\ref{CSC}), and not only when it is studied in the mean-field approximation.
Thereby, we show that the full TDP $\Omega(\Sigma,\Pi,\Delta)$ is invariant under the transformation 
$\widetilde{\cal D}_1$ (\ref{9}). 

To verify the validity of this statement, let us first prove the invariance of the Lagrangian (\ref{100}) under 
the dual transformation ${\cal D}_1$ (\ref{20}). Indeed, as it is clear from the previous subsection, the first term 
$L_{NJL}$ on the right side of Eq. (\ref{100}) is invariant under the action of ${\cal D}_1$. The invariance 
of the second term is the result of the following simple transformations (see also Eqs. (\ref{10})-(\ref{12})),
\begin{eqnarray}
\Delta_A(x)\sim\overline{\psi^{c}}i \gamma^{5}\tau_{2}\lambda_A \psi&=&\psi^{T}Ci \gamma^{5}\tau_{2}\lambda_A \psi=
\psi_{R}^{T}Ci \tau_{2}\lambda_A \psi_{R}
-\psi_{L}^{T}C i\tau_{2}\lambda_A\psi_{L}\nonumber\\
&\stackrel{{\cal D}_1}{\longrightarrow}& \psi_{R}^{T}Ci \tau_{2}\lambda_A \psi_{R}-\psi_{L}^{T}i\tau_{1}C
i\tau_{2}\lambda_A i\tau_{1}\psi_{L}\nonumber\\
&=&\psi_{R}^{T}Ci \tau_{2}\lambda_A \psi_{R}-\psi_{L}^{T}Ci\tau_{2}\lambda_A\psi_{L}=\overline{\psi^{c}}i
\gamma^{5}\tau_{2}\lambda_A \psi. \label{300}
\end{eqnarray}
In addition, it is also necessary to take into account the validity of a similar relation,
\begin{eqnarray}
\Delta_A^*(x)\sim\overline\psi i\gamma^{5} \tau_{2}\lambda_A\psi^c
&\stackrel{{\cal D}_1}{\longrightarrow}\overline\psi i\gamma^{5} \tau_{2}\lambda_A\psi^c.\label{310}
\end{eqnarray}
Hence, $L_{CSC}$ is invariant under the ${\cal D}_1$ duality transformation (\ref{20}). 

Moreover, it is easy to see that the  full TDP $\Omega(\Sigma,\Pi,\Delta)$ of the model 
is also invariant with respect to the dual transformation ${\cal D}_1$. However, it is clear that
in terms of the quantities 
$\Sigma,\Pi$ and $\Delta$ the dual ${\cal D}_1$ transformation looks like $\widetilde{\cal D}_1$ (\ref{9}) (with
unchanged value of $\Delta$, in addition). So the full TDP $\Omega(\Sigma,\Pi,\Delta)$ remains invariant under the action of
$\widetilde{\cal D}_1$. As a consequence, in quark matter which is described by the microscopic Lagrangian
(\ref{100})-(\ref{CSC}) there is a duality between CSB and charged PC phenomena. %\vspace{0.2cm} 

$\bullet$ In conclusion, let us pay attention to the following interesting fact established in this section: two formally different massless NJL models (\ref{1}) and (\ref{100}), which nonetheless effectively describe the physical processes of QCD in the corresponding low energy and density regions, have dual symmetry ${\cal D}_1$ (\ref{20}) between CSB and chargrd PC phenomena. From this, the conclusion suggests itself that the Lagrangian of the massless QCD must also be invariant under the dual transformation ${\cal D}_1$. But this property of the QCD Lagrangian will be considered in more detail below in Section 
IV.

\section{Dual symmetries of dense two-color NJL Lagrangian }

It is well known that in the chiral limit the two-color QCD with $u$ and $d$ quarks is symmetrical not only with respect to the usual flavor $SU(2)_L\times SU(2)_R$ group, but also with an enlarged flavor $SU(4)$ transformation group of quark fields \cite{kogut,son2}. In two-color QCD the colorless baryons are formed by two quarks, i.e. baryons in this theory are bosons. So it is natural to assume that in this theory $u$ and $d$ quarks have a baryon charge equal to 1/2, and their electric charges are 1/2 and (-1/2), correspondingly \cite{ramos}.

To obtain an effective 4-fermion NJL model that would reproduce the basic low-energy properties of the initial two-color QCD theory and have the same symmetry group, the authors of Ref. \cite{weise} integrated out the gluon fields in the Green function generated functional of the QCD. Then, after ``approximating''
the nonperturbative gluon propagator by a delta-function, one arrives at an effective local chiral four-quark interaction of the form (color current)$\times$(color current) of the NJL type describing low-energy hadron physics. Finally, by performing a Fierz transformation of this interaction term and taking into account only scalar and pseudo-scalar $(\overline\psi\psi)$-
as well as scalar $(\psi\psi)$-interaction channels, one obtains a four-fermionic
model given by the following Lagrangian (in Minkowski space-time
notation) \footnote{The most general Fierz transformed four-fermion
interaction includes additional vector and axial-vector $(\overline\psi\psi)$ as well as pseudo-scalar, vector and axial-vector-like $(\psi\psi)$-interactions. However, these terms are omitted
here for simplicity.}
\begin{eqnarray}
L=\overline \psi \Big [i\gamma^\rho\partial_\rho-m_0\Big ]\psi+H\Big [(\overline \psi\psi)^2+(\overline \psi i\gamma^5\vec\tau \psi)^2+
\big (\overline \psi i\gamma^5\sigma_2\tau_2\psi^c\big )\big (\overline{\psi^c}i\gamma^5\sigma_2\tau_2 \psi\big )
\Big]. \label{22}
\end{eqnarray}
In (\ref{22}), 
$\psi^c=C\overline\psi^T$, $\overline{\psi^c}=\psi^T C$ are charge-conjugated spinors, and $C=i\gamma^2\gamma^0$ is
the charge conjugation matrix; $\sigma_2$ is the usual Pauli
matrix acting in the two-dimensional color space. Other symbols are the same as in Eq. (\ref{1}).
 The Lagrangian
(\ref{22}) is invariant with respect to the color $SU(2)_c$ and baryon $U(1)_B$ symmetries. Moreover, at $m_0=0$, it has
the same flavor $SU(4)$ symmetry as two-color QCD (in the literature, the name Pauli-Gursey symmetry is often used for it).

Generally speaking, Lagrangian (\ref{22}) describes physical processes in a vacuum. In order to study the physics of dense two-color quark matter, it is necessary to add in Eq. (\ref{22}) several terms with chemical potentials,
\begin{eqnarray}
L\longrightarrow L_{NJL_2}=L+\overline\psi \gamma^0{\cal M}\psi\equiv L+\overline\psi \gamma^0
\left[\frac{\mu_B}{2}+\frac{\mu_I}2\tau_3
+\frac{\mu_{I5}}2\gamma^5\tau_3+\mu_5\gamma^5\right]\psi. \label{23}
\end{eqnarray}
But in this case even at $m_0=0$ the Lagrangian (\ref{23}), due to the terms with chemical potentials, is no longer invariant with respect to flavor $SU(4)$ symmetry. Instead, in the chiral limit the Lagrangian (\ref{23}), apart from the color $SU(2)_c$, is invariant with respect to the
$U(1)_B\times U(1)_{I_3}\times U(1)_{AI_3}$ transformation group, where $\{U(1)_B:\psi\to\exp (\mathrm{i}\alpha/2) \psi\}$, and abelian isospin $U(1)_{I_3}$ and axial isospin $U(1)_{AI_3}$ transformations are presented in Eq. (\ref{2}).
Below we establish some dual properties of the massless, $m_0=0$, NJL Lagrangian (\ref{23}).

To obtain the phase structure of the two-color NJL model (\ref{23}) at $m_0=0$, it is convenient, as in the three-color case, to use the auxiliary Lagrangian,
\begin{eqnarray}
\widetilde{L}_{NJL_2}=\overline\psi \Big [i\gamma^\rho\partial_\rho+\gamma^0{\cal M} -\sigma -i\gamma^5\vec\tau\vec\pi\Big ]\psi-\frac{\sigma^2+\vec\pi^2+
\delta^*\delta}{4H}-\frac{\delta}{2}\Big [\overline\psi i\gamma^5\sigma_2\tau_2\psi^c\Big ]-
\frac{\delta^*}{2}\Big [\overline{\psi^c}i\gamma^5\sigma_2\tau_2\psi\Big ], \label{24}
\end{eqnarray}
where ${\cal M}$ is introduced in Eq. (\ref{23}) and $\sigma\equiv\sigma (x)$,
$\vec\pi\equiv\vec\pi (x)$, $\delta\equiv\delta (x)$ and $\delta^*\equiv\delta^* (x)$ are auxiliary bosonic fields. %Clearly, the Lagrangians (\ref{23}) and (\ref{24}) are equivalent at $m_0=0$, as can be seen by using the Euler-Lagrange equations of motion for these auxiliary bosonic fields, which take the form

On the basis of the auxiliary Lagrangian (\ref{24}) it is possible to talk about the full TDP 
$\Omega(\sigma,\vec\pi,\delta,\delta^*)$ of the two-color massless NJL model (\ref{23})-(\ref{24}) 
(see the corresponding discussion between (\ref{5}) and (\ref{7})), which supplies us with phase structure of 
dense quark matter composed of massless two-color quarks. Indeed, if in the GMP of the TDP vs 
$\sigma,\vec\pi,\delta,\delta^*$ only $\sigma$ and/or $\pi_0$ are nonzero, then CSB phase is realized. 
If $\pi_1\ne 0$ and/or $\pi_2\ne 0$, the charged PC is observed. Finally, if it is proved that in the GMP we have 
$\delta\ne 0$, then in the system spontaneous breaking of the baryon $U(1)_B$ symmetry occurs, and the baryon 
superfluid (BSF) phase is realized in the model. Fortunately, in our case (as in the case of the three-color massless NJL 
model (\ref{1})), the task of finding the global minimum point of the TDP is greatly simplified, if we take into 
account that the full TDP $\Omega(\sigma,\vec\pi,\delta,\delta^*)$ is
really invariant with respect to the $U(1)_B\times U(1)_{I_3}\times U(1)_{AI_3}$ group. In this case the TDP
of the model is a function of only three field combinations, $\Sigma=\sqrt{\sigma^2+\pi_0^2}$, $\Pi=\sqrt{\pi_1^2+\pi_2^2}$ and $\Delta=\sqrt{\delta\delta^*}$, i.e.
\begin{eqnarray}
\Omega(\sigma,\vec\pi,\delta,\delta^*)\equiv\Omega(\Sigma,\Pi,\Delta).\label{26}
\end{eqnarray}
As a consequence, if the TDP (\ref{26}) has the GMP of the form $(\Sigma\ne 0,\Pi=0,\Delta=0)$, or
$(\Sigma=0,\Pi\ne 0,\Delta=0)$ or $(\Sigma= 0,\Pi=0,\Delta\ne 0)$, then the CSB, or charged PC, or BSF phase is 
observed, respectively, in dense baryon matter composed of two-color quarks. Recall that in Refs. 
\cite{kkz20,u} the phase structure of such an exotic quark matter has been investigated in the framework of the NJL model (\ref{22})-(\ref{23}) at $m_0=0$ in the mean-field approximation. It turns out that in this case the thermodynamic potential $\Omega_{mf}(\Sigma,\Pi,\Delta)$ of the model is invariant under the dual transformation $\widetilde{\cal D}_1$ (\ref{9}) and, in addition, under the following two new discrete dual transformations,
\footnote{Strictly speaking, in Ref. \cite{kkz20} the phase structure of the massless model (\ref{22}) at
$\mu_B\ne 0$, $\mu_I\ne 0$, $\mu_{I5}\ne 0$, but at $\mu_5=0$, was studied. However, in a more recent paper 
\cite{u} the fact was established that dual symmetries (\ref{9}), (\ref{27}) and (\ref{29}) of the 
mean-field TDP of the two-color massless NJL model (\ref{22}) are also presented when all four chemical potentials 
are nonzero.}
\begin{eqnarray}
\widetilde{\cal D}_2:&&~~\Delta \longleftrightarrow\Pi,~~~\mu_{I} \longleftrightarrow \mu_{B},\label{27}\\
\widetilde{\cal D}_3:&&~~\Sigma \longleftrightarrow\Delta,~~~\mu_{B} \longleftrightarrow \mu_{I5},\label{29}
\end{eqnarray}
which lead to the dualities between CSB and charged PC, between charged PC and BSF, and also between CSB and BSF phenomena,
respectively, in dense quark medium with isospin and chiral asymmetries. This greatly simplifies the study of the
phase structure of dense matter, etc. (A list of some additional opportunities for studying these phenomena of dense
medium, which appear due to the dual symmetries of the thermodynamic potential, is presented, e.g., in the paragraphs
following formula (\ref{9}) and in Refs. \cite{kkz20,u}.) The main purpose of this section is to show that the dual symmetries $\widetilde{\cal D}_1$, $\widetilde{\cal D}_2$ and $\widetilde{\cal D}_3$ are not something characteristic 
only for the mean-field approximation of the TDP (\ref{26}). There are also corresponding dual symmetries of the initial Lagrangian 
(\ref{23})-(\ref{24}), due to which we can draw a conclusion about the invariance of the complete TDP (\ref{26}) of the model with respect to the dual 
transformations (\ref{9}), (\ref{27}) and (\ref{29}).

$\bullet$ First of all let us show that under the $PG_1$ transformation (\ref{10})-(\ref{12}) the diquark
structures of the Lagrangians (\ref{22}) and (\ref{24}) as well as the auxiliary boson fields $\delta (x)$ and $\delta^*(x)$ remain intact. Indeed, since $\psi_R^T=\psi^T\Pi_+$ and $\psi_L^T=\psi^T\Pi_-$, we have from Eq. (\ref{10}) 
\begin{eqnarray}
\overline{\psi^{c}} \sigma_{2}\tau_{2}\gamma^{5} \psi&=&\psi^{T}C \sigma_{2}\tau_{2}\gamma^{5} \psi=\psi_{R}^{T}C \sigma_{2}\tau_{2} \psi_{R}
-\psi_{L}^{T}C \sigma_{2}\tau_{2}\psi_{L}\nonumber\\
&\stackrel{PG_1}{\longrightarrow}& \psi_{R}^{T}C \sigma_{2}\tau_{2} \psi_{R}-\psi_{L}^{T}i\tau_{1}C
\sigma_{2}\tau_{2}i\tau_{1}\psi_{L}\nonumber\\
&=&\psi_{R}^{T}C \sigma_{2}\tau_{2} \psi_{R}-\psi_{L}^{T}C \sigma_{2}\tau_{2}\psi_{L}=\overline{\psi^{c}} \sigma_{2}\tau_{2}\gamma^{5} \psi, \label{30}
\\%\end{eqnarray}
%\begin{eqnarray}
\overline\psi \sigma_{2}\tau_{2}\gamma^{5} \psi^c&=&\overline{\psi_L} \sigma_{2}\tau_{2}C\overline{\psi_L}^T-\overline{\psi_R} \sigma_{2}\tau_{2}C\overline{\psi_R}^T\nonumber\\
&\stackrel{PG_1}{\longrightarrow}&
\overline{\psi_L}(-i)\tau_1 \sigma_{2}\tau_{2}C(-i)\tau_1\overline{\psi_L}^T-\overline{\psi_R} \sigma_{2}\tau_{2}C\overline{\psi_R}^T\nonumber\\
&=&\overline{\psi_L} \sigma_{2}\tau_{2}C\overline{\psi_L}^T-\overline{\psi_R} \sigma_{2}\tau_{2}C\overline{\psi_R}^T=\overline\psi \sigma_{2}\tau_{2}\gamma^{5} \psi^c,\label{31}
\end{eqnarray} 
Now, taking into account how other terms of these Lagrangians are transformed under the action of $PG_1$
(see in the previous section), one can conclude that the NJL Lagrangian (\ref{23}) at $m_0=0$ is invariant under the action of the discrete duality transformation ${\cal D}_1$ (\ref{20}). 
This leads to the fact that the total TDP $\Omega(\Sigma,\Pi,\Delta)$ (\ref{26}) of the model, and not only its 
mean-field approximation $\Omega_{mf}(\Sigma,\Pi,\Delta)$, is also symmetric with respect to the dual ${\cal D}_1$ 
transformation. It can be shown in a similar way as in the case with three-color NJL model of the section II. 
Moreover, it is evident that when acting on the full thermodynamic potential $\Omega(\Sigma,\Pi,\Delta)$, the dual 
transformation ${\cal D}_1$ looks like the $\widetilde{\cal D}_1$ transformation (\ref{9}). Thus, the duality between 
the CSB and charged PC phenomena is inherent to the full phase portrait of the model (\ref{23}) as a whole, and not 
only to its mean field approximation. % $\Omega_{mf}(\Sigma,\Pi,\Delta)$.

$\bullet$ Let us introduce the following $PG_2$  transformation of spinor fields (as will soon become clear, the $PG_2$ is actually a Pauli--Gursey-type transformation)
\begin{eqnarray}
PG_2:~~\psi_u\to\psi_u,~~~\psi_d\to \sigma_2\psi_d^c, ~~~{\rm i.e.}~~~PG_2\psi= PG_2\left( {\begin{array}{c}
	 \psi_u \\
	 \psi_d
	\end{array} } \right)=\left( {\begin{array}{c}
	 \psi_u \\
	 \sigma_2\psi_d^c
	\end{array} } \right),\label{32}
\end{eqnarray}
where $\sigma_2$ is the Pauli matrix acting in the two-dimensional color space. Then it is easy to check that
\begin{eqnarray}
\overline\psi\psi&=&\overline\psi_u\psi_u+\overline\psi_d\psi_d\stackrel{PG_2}{\longrightarrow}
\overline\psi_u\psi_u+\overline{(\sigma_2\psi^c_d)}(\sigma_2\psi^c_d)=
\overline\psi_u\psi_u+\overline{\psi^c_d}\psi^c_d=\overline\psi\psi,\nonumber\\
\overline\psi\gamma^5\tau_3\psi&=&\overline\psi_u\gamma^5\psi_u-\overline\psi_d\gamma^5\psi_d
\stackrel{PG_2}{\longrightarrow} \overline\psi_u\gamma^5\psi_u-\overline{(\sigma_2\psi^c_d)}
\gamma^5(\sigma_2\psi^c_d)= \overline\psi_u\gamma^5\psi_u-\overline{\psi^c_d}\gamma^5\psi^c_d=\overline\psi\gamma^5\tau_3\psi,\nonumber\\
\overline\psi\gamma^0\psi&=&\overline\psi_u\gamma^0\psi_u+\overline\psi_d\gamma^0\psi_d
\stackrel{PG_2}{\longrightarrow}\overline\psi_u\gamma^0\psi_u+\overline{(\sigma_2\psi^c_d)}\gamma^0
(\sigma_2\psi^c_d)=\overline\psi_u\gamma^0\psi_u+\overline{\psi^c_d}\gamma^0\psi^c_d\nonumber\\
&=&\overline\psi_u\gamma^0\psi_u-\overline{\psi_d}\gamma^0\psi_d=\overline\psi\gamma^0\tau_3\psi,\nonumber\\
\overline\psi\gamma^0\tau_3\psi&=&\overline\psi_u\gamma^0\psi_u-\overline\psi_d\gamma^0\psi_d
\stackrel{PG_2}{\longrightarrow}\overline\psi_u\gamma^0\psi_u-\overline{(\sigma_2\psi^c_d)}\gamma^0
(\sigma_2\psi^c_d)=\overline\psi_u\gamma^0\psi_u-\overline{\psi^c_d}\gamma^0\psi^c_d\nonumber\\
&=&\overline\psi_u\gamma^0\psi_u+\overline{\psi_d}\gamma^0\psi_d=\overline\psi\gamma^0\psi, %,\nonumber\\&&
\label{33}
\end{eqnarray}
i.e. $\overline\psi\gamma^0\psi$$\stackrel{PG_2}{\longleftrightarrow}$$\overline\psi\gamma^0\tau_3\psi$.
Moreover, it is possible to show that under the action of $PG_2$ the chiral, $\bar\psi\gamma^0\gamma^5\psi$, and the chiral isospin, $\bar\psi\gamma^0\gamma^5\tau_3\psi$, density operators are invariant. Let us consider the following four-fermion structures responsible for the charged PC and BSF interaction channels, respectively,
\begin{eqnarray}
\big (4F\big )_{pc}&\equiv&(i\overline\psi\gamma^5\tau_1\psi)^2+(i\overline\psi\gamma^5\tau_2\psi)^2=
-\Big [\overline\psi_u\gamma^5\psi_d+\overline\psi_d\gamma^5\psi_u\Big]^2+\Big [\overline\psi_u\gamma^5\psi_d-\overline\psi_d\gamma^5\psi_u\Big]^2=-4
(\overline\psi_u\gamma^5\psi_d)(\overline\psi_d\gamma^5\psi_u),\nonumber\\
\big (4F\big )_{bsf}&\equiv&\big (\overline\psi i\gamma^5\sigma_2\tau_2\psi^c\big )\big (\overline{\psi^c}i\gamma^5\sigma_2\tau_2\psi\big )\nonumber\\
&=&\Big [\overline\psi_u\sigma_2\gamma^5\psi_d^c-\overline\psi_d\sigma_2\gamma^5\psi_u^c\Big]\Big [\overline{\psi_u^c}\sigma_2\gamma^5\psi_d-
\overline{\psi_d^c}\sigma_2\gamma^5\psi_u\Big]=-4\big(\overline\psi_u\sigma_2\gamma^5\psi_d^c\big)
\big(\overline{\psi_d^c}\sigma_2\gamma^5\psi_u\big).
\label{34}
\end{eqnarray}
To obtain the last expression in Eq. (\ref{34}), we use in square brackets there two rather general relations,
\begin{eqnarray}
\overline{\chi^c}U \eta&=&-\overline{\eta^c}\big(CUC\big)^T \chi,~~\overline\psi V \zeta^c=-\overline\zeta\big(CVC\big)^T \psi^c, \label{35}
\end{eqnarray}
in which $\chi,\eta,\psi$ and $\zeta$ are Dirac spinors, $U$ and $V$ are arbitrary matrices in spinor space, and the charge conjugation matrix $C$ is define below Eq. (\ref{22}).
Since under the transformation $PG_2$ we have $\sigma_2\psi_d^c\rightarrow -\psi_d$ and
$\overline{\psi_d^c}\sigma_2\rightarrow -\overline\psi_d$, it follows from  Eqs. (\ref{34}) that
$\big (4F\big )_{pc}\stackrel{PG_2}{\longleftrightarrow}\big (4F\big )_{bsf}$, i.e. $PG_2$ is indeed a 
Pauli--Gursey-type transformation. In addition, it is clear from Eq. (\ref{33}) that the four-fermion term $(\overline\psi\psi)^2+(i\overline\psi\gamma^5\tau_3\psi)^2$, responsible for the CSB interaction channel,
remains intact by the action of the $PG_2$.
Since the free Dirac Lagrangian $\overline\psi \big (i\gamma^\rho\partial_\rho-m_0\big )\psi$ is also invariant under the action of $PG_2$ (the invariance of its massive and kinetic terms are shown in Eq. (\ref{33}) and below in Eq. (\ref{48}), respectively), we see from the above relations that when the
discrete $PG_2$ transformation (\ref{32}) acts on spinor fields and simultaneously the permutation of the chemical 
potentials $\mu_B$ and $\mu_I$ occurs, i.e. the following ${\cal D}_2$ transformation is performed in the system,
\begin{eqnarray}
{\cal D}_2:~~\psi_u\to\psi_u,~~\psi_d\longrightarrow \sigma_2\psi_d^c,~~\mu_B\longleftrightarrow \mu_I,\label{36}
\end{eqnarray}
then the two-color NJL Lagrangian (\ref{23}) remains invariant. \footnote{\label{p3} Recall, in the present section we study properties of the massless NJL model (\ref{23}) but, as it is clear from above consideration, the dual symmetry ${\cal D}_2$ is also realized in this model even at nonzero values of $m_0$.}
Furthermore, it means that under the dual ${\cal D}_2$ transformation the full TDP 
$\Omega(\Sigma,\Pi,\Delta)$ (\ref{26}) of the massless model is also invariant. One can understand how the transformation ${\cal D}_2$ looks like in terms of the variables $\Sigma$, $\Pi$ and $\Delta$, it is clear that $\Pi\leftrightarrow\Delta$ under the 
$PG_2$ (or ${\cal D}_2$ (\ref{36})) transformation, but $\Sigma$ remains intact. Hence, we see that the ${\cal D}_2$ transformation 
(\ref{36}) is indeed a dual symmetry of the model, which moves the BSF and charged PC channels into each other. So in terms of the full TDP 
$\Omega(\Sigma,\Pi,\Delta)$ (\ref{26}) of the model the dual symmetry ${\cal D}_2$ looks like the 
$\widetilde{\cal D}_2$ dual symmetry (\ref{27}) of the mean-field TDP of the two-color NJL model. Hence, the duality 
between BSF and charged PC phenomena is inherent to 
the full phase portrait of the model (\ref{23}) as well, and not only to its mean field approximation.

$\bullet$ Now, we want to find such a (dual) transformation ${\cal D}_3$ of the field variables and chemical 
potentials of the massless Lagrangian (\ref{23}), under which it is invariant, and such phenomena of dense baryonic
medium as CSB and BSF are dually symmetric (conjugated), i.e. go one into another in its full phase diagram. It is
obvious that in terms of the thermodynamic potential $\Omega(\Sigma,\Pi,\Delta)$ (\ref{26}) the ${\cal D}_3$ should 
look like the dual transformation $\widetilde{\cal D}_3$ of Eq. (\ref{29}) for its mean-field approximation 
$\Omega_{mf}(\Sigma,\Pi,\Delta)$. Since it is quite easy to notice that $\widetilde{\cal D}_3=\widetilde{\cal D}_1
\cdot\widetilde{\cal D}_2\cdot\widetilde{\cal D}_1$, it is natural to look for the corresponding Pauli--Gursey-type
transformation $PG_3$ of flavor doublet $\psi(x)$ field in the form
\begin{eqnarray}
PG_3=PG_1\cdot PG_2\cdot PG_1, \label{380}
\end{eqnarray}
where $PG_1$ and $PG_2$ are Pauli--Gursey-type transformations presented in Eqs. (\ref{12}) and (\ref{32}), 
respectively. So it is easy to find that
\begin{eqnarray}
PG_3:~~\psi\equiv \left( {\begin{array}{c}
	 \psi_{u} \\
	 \psi_d
	\end{array} } \right)\longrightarrow\left( {\begin{array}{c}
	 \psi_{uR}+i\sigma_2\big (\psi_d^c\big )_L \\
	 i\sigma_2\big (\psi_u^c\big )_R-\psi_{dL}
	\end{array} } \right),\label{38}
\end{eqnarray}
where we have used the relations valid for arbitrary four-component spinor $\chi (x)$,
\begin{equation}
\big (\chi_R\big )^c=\big (\chi^c\big )_L,~~~~~~\big (\chi_L\big )^c=\big (\chi^c\big )_R.\label{39}
\end{equation}
The relation (\ref{38}) means that
\begin{equation}
\psi_u\stackrel{PG_3}{\longrightarrow}\psi_{uR}+i\sigma_2\big (\psi_d^c\big )_L,~~
\psi_d\stackrel{PG_3}{\longrightarrow}
i\sigma_2\big (\psi_u^c\big )_R-\psi_{dL}.\label{40}
\end{equation}
Since $PG_3$ is the product of three transformations acting one after another, each of which leaves the Lagrangian (\ref{22}) invariant at zero bare mass $m_0$, then this quantity is also symmetric with respect to $PG_3$.
To understand how the chemical potential terms of the Lagrangian (\ref{23}) are transformed under $PG_3$, let us first
recall that the chiral density operator $\overline\psi\gamma^0\gamma^5\psi$ is invariant both under $PG_1$
and $PG_2$ field transformations (see the remarks before Eq. (\ref{18}) and just after Eq. (\ref{33})). Hence, the
$PG_3$ transformation, due to its structure (\ref{380}), does not change this quantity. Second, since
$\overline\psi\gamma^0\tau_3\psi\stackrel{PG_1}{\longleftrightarrow}\overline\psi\gamma^0\gamma^5\tau_3\psi$
but $\overline\psi\gamma^0\gamma^5\tau_3\psi$ remains intact under the $PG_2$, one can conclude that $PG_3$ does not
change the isospin density operator $\overline\psi\gamma^0\tau_3\psi$. However the quark number
$\overline\psi\gamma^0\psi$ and chiral isospin $\overline\psi\gamma^0\gamma^5\tau_3\psi$ density operators turn into
each other under the action of $PG_3$. Indeed, as it is clear from Eqs. (\ref{18}) and (\ref{33}), the following relations are valid
\begin{equation}
\overline\psi\gamma^0\psi\stackrel{PG_1}{\longleftrightarrow}\overline\psi\gamma^0\psi
\stackrel{PG_2}{\longleftrightarrow}\overline\psi\gamma^0\tau_3\psi\stackrel{PG_1}{\longleftrightarrow}
\overline\psi\gamma^0\gamma^5\tau_3\psi,\label{41}
\end{equation}
i.e. we have
\begin{equation}
\overline\psi\gamma^0\psi\stackrel{PG_3}{\longleftrightarrow}\overline\psi\gamma^0\gamma^5\tau_3\psi.
\label{42}
\end{equation}
Alternatively, the relation (\ref{42}) can be obtained by direct use of the definition (\ref{40}) for $PG_3$.
For example, applying the $PG_3$ (\ref{40}) to the quark number density $\overline\psi\gamma^0\psi$, we have
\begin{eqnarray}
\overline\psi\gamma^0\psi&=&\overline\psi_u\gamma^0\psi_u+\overline\psi_d\gamma^0\psi_d=
\overline{\psi_{uR}}\gamma^0\psi_{uR}+\overline{\psi_{uL}}
\gamma^0\psi_{uL}+\overline{\psi_{dR}}\gamma^0\psi_{dR}+\overline{\psi_{dL}}\gamma^0\psi_{dL}\nonumber\\
&\stackrel{PG_3}{\longrightarrow}&\overline{\big (\psi_{uR}+i\sigma_2\big (\psi_d^c\big )_L\big )}\gamma^0
\big (\psi_{uR}+i\sigma_2\big (\psi_d^c\big )_L\big )+\overline{\big (i\sigma_2\big (\psi_u^c\big )_R-\psi_{dL}
\big )}\gamma^0
\big (i\sigma_2\big (\psi_u^c\big )_R-\psi_{dL}\big )\nonumber\\
&=&\overline{\psi_{uR}}\gamma^0\psi_{uR}+\overline{\big (\psi_d^c\big )_L}\gamma^0\big (\psi_d^c\big )_L+
\overline{\big (\psi_u^c\big )_R}\gamma^0\big (\psi_u^c\big )_R+\overline{\psi_{dL}}\gamma^0\psi_{dL}\nonumber\\
&=&\overline{\psi_{uR}}\gamma^0\psi_{uR}-\overline{\psi_{dR}}\gamma^0\psi_{dR}-
\overline{\psi_{uL}}\gamma^0\psi_{uL}+\overline{\psi_{dL}}\gamma^0\psi_{dL}=
\overline\psi\gamma^0\gamma^5\tau_3\psi.\label{43}
\end{eqnarray}
In deriving the relations (\ref{43}) we use Eq. (\ref{39}) as well as the well-known relationship
$\bar\chi^c\gamma^0\chi^c=-\bar\chi\gamma^0\chi$. Hence, the NJL two-color Lagrangian (\ref{23}) at $m_0=0$ is
invariant under simultaneous action of two transformations, the $PG_3$  of Eq. (\ref{40}) and $\mu_B\longleftrightarrow\mu_{I5}$. We
call it the dual ${\cal D}_3$ transformation, i.e.
\begin{equation}
{\cal D}_3:~~\psi_u\longrightarrow\psi_{uR}+i\sigma_2\big (\psi_d^c\big )_L,~~
\psi_d\longrightarrow
i\sigma_2\big (\psi_u^c\big )_R-\psi_{dL},~~\mu_B\longleftrightarrow\mu_{I5}.\label{44}
\end{equation}
(It follows from Eq. (\ref{380}) that ${\cal D}_3={\cal D}_1\cdot {\cal D}_2\cdot {\cal D}_1$.)
As a consequence, it is clear that the full thermodynamic potential $\Omega(\Sigma,\Pi,\Delta)$ (\ref{26}) of this 
model is also symmetric with respect to the dual transformation ${\cal D}_3$. Moreover, the transformation 
${\cal D}_3$ in terms of the total TDP $\Omega(\Sigma,\Pi,\Delta)$ looks like the dual transformation $\widetilde{\cal D}_3$ (\ref{29}) of 
its mean-field approximation $\Omega_{mf}(\Sigma,\Pi,\Delta)$.
This conclusion can be easily drawn knowing the following transformations of the 
four-fermionic structures (\ref{34})
under the $PG_1$ and $PG_2$ (see in Eq. (\ref{17}) and the remark below Eq. (\ref{35})),
\begin{eqnarray}
\big (4F\big )_{csb}\stackrel{PG_1}{\longleftrightarrow}\big (4F\big )_{pc}&\stackrel{PG_2}{\longleftrightarrow}&\big (4F\big )_{bsf}\stackrel{PG_1}{\longleftrightarrow}\big (4F\big )_{bsf},\nonumber\\
\big (4F\big )_{pc}\stackrel{PG_1}{\longleftrightarrow}\big (4F\big )_{csb}&\stackrel{PG_2}{\longleftrightarrow}&\big (4F\big )_{csb}\stackrel{PG_1}{\longleftrightarrow}\big (4F\big )_{pc}.
\label{45}
\end{eqnarray}
Indeed, since $PG_3=PG_1\cdot PG_2\cdot PG_1$ then it is evident from Eq. (\ref{45}) that 
$\big (4F\big )_{csb}\stackrel{PG_3}{\longleftrightarrow}\big (4F\big )_{bsf}$ and
$\big (4F\big )_{pc}\stackrel{PG_3}{\longleftrightarrow}\big (4F\big )_{pc}$. Hence, the order parameters $\Sigma$ and $\Delta$ of the CSB and BSF phases are transformed into each other by $PG_3$ as well as by ${\cal D}_3$,  $\Sigma\longleftrightarrow\Delta$ (the order parameter $\Pi$ remains intact in this case). 

So the duality between CSB and BSF phenomena in dense quark matter with isospin and chiral imbalances is an exact 
property of the full phase diagram of two-color massless NJL model (\ref{23}), and not only of its mean-field 
approximation, as it was shown in Ref. \cite{kkz20}.

\section{Dual symmetries of dense two- and three-color QCD }
\subsection{Invariance of Lagrangian}
It is well known that NJL models (\ref{1}) and (\ref{100}) describe effectively different low-energy 
regions of QCD with three-color quarks, whereas (\ref{23}) is the low-energy effective model of the two-color QCD. 
As a consequence, these (we still consider only models with massless quarks) NJL models are invariant with respect to the 
same symmetry groups as the original quantum chromodynamics, i.e., apart from $SU(3)_c$ and $SU(2)_c$ 
color groups, respectively, at nonzero chemical potentials the models (\ref{1}), (\ref{100}) and (\ref{23}) are 
invariant under the transformations from 
$U(1)_B\times U(1)_{I_3}\times U(1)_{AI_3}$ group (see Eq. (\ref{2}) and remarks after Eq. 
(\ref{23})). However, basing on the results of previous sections, we see that all above mentioned NJL Lagrangians 
%(\ref{1}) and (\ref{23}) 
are invariant in the chiral limit under the so-called duality transformation 
${\cal D}_1$ (\ref{20}) in addition. Moreover,
the two-color NJL Lagrangian (\ref{23}) is symmetric with respect to two additional dual transformations, ${\cal D}_2$ 
(\ref{36}) and ${\cal D}_3$ (\ref{44}). Does it mean that the original massless QCD theories have the same duality 
properties as the corresponding low-energy massless NJL models? In this section, we show that this is the case.

Our starting point is the Lagrangian for the quark sector of two- and three-color massless QCD extended by the baryon
$\mu_B$-, isospin $\mu_I$-, chiral $\mu_5$- and chiral isospin $\mu_{I5}$ chemical potentials,
\begin{equation}
L_{QCD}=i\overline\psi \gamma^{\nu}\nabla_{\nu} \psi+\mu \overline{\psi}\gamma^{0}\psi+\mu_5 \overline{\psi}\gamma^{0}
\gamma^5\psi+\nu\overline{\psi}\gamma^{0}\tau_{3}\psi+\nu_{5} \overline{\psi}\gamma^{0}\gamma^{5}\tau_{3}\psi,\label{46}
\end{equation}
where, as in the previous sections, $\psi (x)$ is the two-flavor quark doublet, $\nu=\mu_I/2$, $\nu_5=\mu_{I5}/2$ and 
$\mu=\mu_B/3$ for the case of three-color QCD or $\mu=\mu_B/2$ in the case of two-color QCD. Moreover, symbol 
$\nabla_{\nu}$ in this expression means the color $SU(2)_c$ covariant derivative, i.e. 
$\nabla_{\nu}=\partial_\nu-ig\sigma_aA^a_\nu(x)$, where $\sigma_a$
($a=1,2,3$) are the 2$\times$2 Pauli matrices (for the case of two-color QCD). Or it is the color $SU(3)_c$ 
covariant derivative, $\nabla_{\nu}=\partial_\nu-ig\lambda_aA^a_\nu(x)$,
where $\lambda_a$ ($a=1,..,8$) are the 3$\times$3 Gell-Mann matrices (for the case of three-color QCD).
In both cases, $A^a_\nu(x)$ is a vector gauge field. The Lagrangian (\ref{46}) describes properties
of dense quark matter with isospin and chiral asymmetries. Here we are going to show that the dual symmetries,
typical for massless NJL models (see in the previous sections) also inherent in the 
corresponding massless QCD Lagrangians (\ref{46}).

$\bullet$ First, let us show that both in the two-color and three-color cases the QCD Lagrangian (\ref{46}) is 
invariant under
the ${\cal D}_1$ dual transformation (\ref{20}). Indeed, it is clear from the studies of the previous two sections that
the sum of all chemical potential terms of this Lagrangian is invariant with respect to ${\cal D}_1$ regardless
of whether $SU(2)_c$ or $SU(3)_c$ color group is used in Eq. (\ref{46}). Moreover, the first term in Eq. (\ref{46}) is also
invariant under the action of the ${\cal D}_1$. This is true, because
\begin{eqnarray}
i\overline\psi \gamma^{\nu}\nabla_{\nu} \psi&=&i\overline\psi_{R} \gamma^{\nu}\nabla_{\nu} \psi_{R}+i\overline\psi_{L}
\gamma^{\nu}\nabla_{\nu} \psi_{L}\stackrel{{\cal D}_1}{\longrightarrow}
i\overline\psi_{R} \gamma^{\nu}\nabla_{\nu} \psi_{R}+i\overline\psi_{L}(-i\tau_{1}) \gamma^{\nu}\nabla_{\nu}i\tau_{1}
\psi_{L}\nonumber\\
&=&i\overline\psi_{R} \gamma^{\nu}\nabla_{\nu} \psi_{R}+i\overline\psi_{L} \gamma^{\nu}\nabla_{\nu} \psi_{L}=
i\overline\psi \gamma^{\nu}\nabla_{\nu} \psi.\label{47}
\end{eqnarray}
Hence, the QCD Lagrangian (\ref{46}) as a whole is symmetric with respect to a dual 
transformation ${\cal D}_1$ (\ref{20}).

$\bullet$ To prove that the two-color QCD Lagrangian (\ref{46}) is invariant under ${\cal D}_2$ and ${\cal D}_3$
dual transformations (\ref{36}) and  (\ref{44}), respectively, one should note first of all that the set of all chemical potential terms is symmetric under ${\cal D}_2$
and ${\cal D}_3$ (this is shown in the previous section). Next, consider how the term $i\overline\psi
\gamma^{\nu}\partial_\nu \psi$ of Eq. (\ref{46}) is transformed under the action of ${\cal D}_2$. Using in this case the expression (\ref{36}) as well as the well-known relations \footnote{In the second line of Eq. (\ref{48}) we use the definition of the charge conjugation of an arbitrary spinor field $\psi$, introduced below Eq. (\ref{22}), as well as the relation $(\gamma^\nu)^T=C\gamma^\nu C$. Here we also take into account the anticommuting of spinor fields, and the relations $\psi^T (\gamma^{\nu})^T\partial_{\nu} \overline \psi^T=-\overline \psi \gamma^{\nu}\partial_{\nu}^T \psi$ and $\partial_{\nu}^T=-\partial_{\nu}$.\label{6} }, we have
\begin{eqnarray}
i\overline\psi \gamma^{\nu}\partial_{\nu} \psi&=&i\overline\psi_{u} \gamma^{\nu}\partial_{\nu} \psi_{u}+i\overline\psi_{d}
\gamma^{\nu}\partial_{\nu} \psi_{d}\stackrel{{\cal D}_2}{\longrightarrow}
i\overline\psi_{u} \gamma^{\nu}\partial_{\nu} \psi_{u}+i\overline{\psi_{d}^c}\sigma_{2} \gamma^{\nu}\partial_{\nu}\sigma_{2}
\psi_{d}^c\nonumber\\
&=&i\overline\psi_{u} \gamma^{\nu}\partial_{\nu} \psi_{u}+i\psi_{d}^TC \gamma^{\nu}\partial_{\nu}C \overline\psi_{d}^T=i\overline\psi_{u} \gamma^{\nu}\partial_{\nu} \psi_{u}-i\overline\psi_{d} \gamma^{\nu}\partial_{\nu}^T\psi_{d}\nonumber\\&=&i\overline\psi_{u} \gamma^{\nu}\partial_{\nu} \psi_{u}+i\overline\psi_{d} \gamma^{\nu}\partial_{\nu}\psi_{d}=
i\overline\psi \gamma^{\nu}\partial_{\nu} \psi.\label{48}
\end{eqnarray}
In a similar way it is possible to show that
\begin{eqnarray}
\overline{\psi_d}\gamma^\nu\sigma_aA^a_\nu\psi_d\stackrel{{\cal D}_2}{\longrightarrow}
\overline{\psi_d^c}\sigma_2\gamma^\nu\sigma_aA^a_\nu\sigma_2
\psi_d^c= -\psi_d^T (\gamma^\nu)^T\sigma_a^TA^a_\nu\overline \psi_d^T= \overline{\psi_d}\gamma^\nu\sigma_aA^a_\nu\psi_d,
\label{49}
\end{eqnarray}
where we also take into account that spinor fields anticommute with each other, as well as the validity of
the relation $\sigma_2\sigma_a\sigma_2=-\sigma_a^T$ ($a=1,2,3$). On the basis of the Eqs. (\ref{48}) and (\ref{49})
one can conclude that the first term of the two-color Lagrangian (\ref{46}), i.e. the term
$i\overline\psi \gamma^{\nu}\nabla_{\nu} \psi$, where $\nabla_{\nu}=\partial_\nu-ig\sigma_aA^a_\nu(x)$, is invariant under
the dual transformation ${\cal D}_2$. As a consequence, the total two-color and two-flavor QCD Lagrangian (\ref{46})
is dually ${\cal D}_2$ symmetric. \footnote{As it is clear from Eq. (\ref{33}), the bare mass term
$m_0\overline\psi\psi$ is invariant under the ${\cal D}_2$. Therefore, not only massless, but also massive two-color
QCD is dually ${\cal D}_2$ invariant.}

Finally, since the two-color massless QCD Lagrangian (\ref{46}) is invariant under ${\cal D}_1$ and ${\cal D}_2$ and
since in this case ${\cal D}_3={\cal D}_1\cdot {\cal D}_2\cdot {\cal D}_1$, it is obvious that the two-color
and two-flavor QCD Lagrangian (\ref{46}) is symmetric under the dual ${\cal D}_3$ transformation as well.

The dual symmetries of two- and three-color massless QCD Lagrangians proved above must be reflected in a quite definite way in the 
symmetry properties of their total TDPs, and hence in their complete phase portraits. As a result, we see that at 
$N_c=3$ the CSB phase in the full $(\mu,\mu_I,\mu_{I5},\mu_5)$-phase portrait should be dually conjugated to 
the charged PC phase. If $N_c=2$, then in the full phase portrait of the two-color QCD this dual symmetry should be 
supplemented by two (dual) symmetries, between the CSB and BSF phases as well as between the charged PC and 
BSF phases. These dual symmetries can be considered as fundamental properties of massless QCD, which should 
manifest themselves in the framework of any approximation to the QCD phase diagram.

\subsection{Invariance of path integration measure }
We have proved in the previous section that Lagrangian of two-color QCD is symmetric with respect to three dual transformations and three-color QCD with respect to one. But what if this would be broken on a quantum level by anomalies which is tantamount to invariance of the integration measure ${\cal D}\psi{\cal D}\bar\psi$ of path integrals with respect to the dual transformations?

To answer the question, let us start with 
two-color QCD. The two-color QCD Lagrangian can be rewritten in the form $i\bar\Psi\gamma^{\mu}D_{\mu}\Psi$ \cite{andersen2}, where $\Psi$ is defined as 
$$
\Psi=\left( {\begin{array}{cc}
\psi_{uL} \\
\psi_{dL} \\
\sigma_{2}(\psi_{uR})^{c} \\
\sigma_{2} (\psi_{dR})^{c}
\end{array} } \right),\;\;\;\;\;\;\;\;\;\;\;\;\;\;\;  \bar\Psi=\Psi^{\dagger}\gamma^{0}=\left( {\begin{array}{cc}
\bar\psi_{uL},\;
\bar\psi_{dL},\;
\overline{(\psi_{uR})^{c}}\sigma_{2},\;
\overline{(\psi_{dR})^{c}}\sigma_{2}
\end{array} } \right).
$$
Supposing that $\Psi$ transforms under a fundamental multiplet of $U(4)$, one can notice 
that Lagrangian of two-color QCD is invariant with respect to $U(4)$ group. But its subgroup 
$U(1)$, composed of transformations $\Psi\to e^{-i\alpha}\Psi$, is broken down by anomaly 
\cite{Smilga:1994tb, andersen2}. Indeed, one can show that this group defined in terms of 
$\Psi$ field is equivalent to more familiar group $U_A(1):\psi\to e^{i\gamma^{5}\alpha}\psi$
(i. e. $\psi_R\to e^{i\alpha}\psi_R$, $\psi_L\to e^{-i\alpha}\psi_L$). So, the symmetry of two-color QCD that is not broken by anomaly is $SU(4)$. 

Now it is interesting to note that the $PG_1$ transformation (\ref{10}) of fermion fields can be rewritten in the 
following form in terms of $\Psi$-field components:
$\psi_1\to e^{i\tau_1 \alpha_1}\psi_1$, where $\psi_1$ is a spinor doublet composed of 
first two components of $\Psi$ spinor, namely $\psi_1=\left( {\begin{array}{c}
\psi_{uL} \\
\psi_{dL}
\end{array} } \right)$,
and $\alpha_1=\pi/2$, while the 
the fields $\psi_{uR}$ 
and $\psi_{dR}$, i.e. the
last two components of $\Psi$, do not transform. 
One can easily see that it is an element of the $SU(2)$ subgroup of a more general $SU(4)$ 
transformation group of the field $\Psi$, $e^{i\tau_1 \alpha_1}\in SU(2)\subset SU(4)$. 
Therefore, it is natural to conclude that the symmetry of the two-color QCD under the $PG_1$ 
fermion field transformation (\ref{10}) is not broken by anomaly. Hence, the fermion 
integration measure $D\psi D\bar\psi$ is invariant with respect to this transformation, 
and duality ${\cal D}_1$ is a symmetry not only of the two-color QCD Lagrangian, but also of 
full quantum theory and, as a result, of full thermodynamic potential, etc.

In a similar way, one can rewrite the original $PG_2$ transformation (\ref{32}) in different form 
in terms of $\Psi$-field components: $\psi_2\to e^{i\tau_2 \alpha_2}\psi_2$, where $\psi_2$ 
is a doublet composed of the second and forth components of the $\Psi$ spinor,
namely $\psi_2=\left( {\begin{array}{c}
 \psi_{dL} \\
\sigma_2(\psi_{dR})^c
\end{array} } \right)$,
and $\alpha_2=\pi/2$, but the fields $\psi_{uL}$ and $\psi_{uR}$, i.e. the second and third components of the $\Psi$ spinor, do not transform.  
One can easily observe that $e^{i\tau_2 \alpha_2}$ is also a transformation from another one $SU(2)$ subgroup of $SU(4)$, i. e. $e^{i\tau_2 \alpha_2}\in SU(2)\subset SU(4)$. Hence, $PG_2$ is anomaly free as well.

One can easily see that since as transformations of $PG_1$ as well as $PG_2$ belongs to $SU(4)$, fermion field transformation of $PG_3$, which is of the form $PG_1\cdot PG_2\cdot PG_1$, belongs to $SU(4)$ group and hence is not broken by anomaly either. 

Now let us turn to three-color QCD. Fermion $PG_1$ field transformation in three-color case also can be rewritten in the form $\left( {\begin{array}{c}
\psi_{uL} \\
\psi_{dL}
\end{array} } \right)\to e^{i\tau_1 \alpha_1}\left( {\begin{array}{c}
\psi_{uL} \\
\psi_{dL}
\end{array} } \right)$, where $\alpha_1=\pi/2$ (right-handed spinors remain intact). Recall that two-flavor and three-color QCD Lagrangian is symmetric with respect to $SU_L(2)\times SU_R(2)\times U(1)\times U_A(1)=SU_V(2)\times SU_A(2)\times U(1)\times U_A(1)$, which is broken by chiral anomaly to $SU_V(2)\times SU_A(2)\times U(1)$. Let us observe that $e^{i\tau_1 \alpha_1}\in SU_L(2)$ or a product of transformations from $SU_V(2)$ and $SU_A(2)$ groups that are not anomalous and hence dual transformation ${\cal D}_1$ is not broken by anomaly. Hence it is a symmetry both of the three-color massless QCD  Lagrangian and fermion integration measure $D\psi D\bar\psi$ in these theory. So full thermodynamic potential is also symmetric with respect to this dual transformation. 

\section{Summary and conclusions}

In the proposed paper, it is shown that the Lagrangians of the massless NJL models (\ref{1}) and (\ref{100}), built from three-color $u$
and $d$ quarks and extended by terms with $\mu_B,\mu_I,\mu_{I5}$ and $\mu_5$ chemical potentials, have the so-called 
dual ${\cal D}_1$ symmetry (\ref{20}) between the CSB and charged PC interaction channels (see in section II). 
It means that in each of these models the total TDP as function of the above chemical potentials and order parameters 
$\Sigma$ and $\Pi$ of the CSB and charged PC phases is symmetric under the transformations (\ref{9}). 
As a result, in the {\it complete} phase diagram of each of these models, the CSB and charged PC phases are 
located dually symmetrically to each other. In other words, at fixed $\mu_B$ and $\mu_5$ on the complete 
$(\mu_I,\mu_{I5})$-phase portrait of each of the models the CSB and charged PC phases should be arranged 
mirror-symmetrically with respect to the $\mu_I=\mu_{I5}$ line. So we have shown 
that dual symmetry between CSB and charged PC phenomena is an inherent property of the massless NJL models 
(\ref{1}) and (\ref{100}) as a whole (and not only of its mean-field approximation, as it was 
established earlier in Refs. \cite{kkz18,u2}).

Further, see in section IV, we prove that the dual symmetry ${\cal D}_1$ (\ref{20}) is also inherent in the more fundamental 
massless QCD models with two- and three-colored $u$ and $d$ quarks. Recall, at zero chemical potentials this QCD theory is 
invariant under $SU(2)_L\times SU(2)_R$ flavor group in the $N_c=3$ case. However, in the case when quarks are two-colored, the flavor 
symmetry group of the massless QCD Lagrangian is extended to the $SU(4)$ group. Due to this fact, the entire massless 
Lagrangian of dense QCD with $N_c=2$ has two additional ${\cal D}_2$ (\ref{36}) and ${\cal D}_3$ (\ref{44}) dual symmetries, 
between the charged PC and BSF and between the CSB and BSF channels, respectively. Therefore, the corresponding 
massless NJL model (\ref{22}), which at low energies is equivalent to the two-color QCD, has the same dual 
symmetries (see in section III). 
It means that ${\cal D}_{1,2,3}$ dual symmetries between different phases are inherent in the phase portrait as a 
whole, both in the massless NJL and QCD models with two-color quarks. (Recall, earlier the (dual) 
symmetries between CSB, charged PC and BSF phases  have been established only in the framework of the 
two-color NJL model, and only in the mean-field approximation \cite{kkz20,u}.)

In conclusion, we should note that  dual symmetries are proved to exist between different phases of dense quark 
matter formed from massless $u$ and $d$ quarks. However, we think that in real situations when quarks are massive, 
traces of dual symmetries of the massless QCD phase portrait should be preserved to some extent. As a confirmation, 
we can refer to the phase portrait calculated in the mean field approximation in the massive NJL model \cite{kkz19}, 
in which the dual symmetry between CSB and charged PC phases is only approximate.


\begin{thebibliography}{999}
\bibitem{thies}
M.~Thies,
%``Duality between quark quark and quark anti-quark pairing in 1+1 dimensional large N models,''
Phys. Rev. D \textbf{68}, 047703 (2003);
Phys. Rev. D \textbf{90}, no.10, 105017 (2014).
%doi:10.1103/PhysRevD.90.105017
%[arXiv:1408.5506 [hep-th]].

\bibitem{ebert}
D.~Ebert, T.~G.~Khunjua, K.~G.~Klimenko and V.~C.~Zhukovsky,
%``Competition and duality correspondence between inhomogeneous fermion-antifermion and fermion-fermion condensations in the NJL$_2$ model,''
Phys. Rev. D \textbf{90}, no.4, 045021 (2014).
%doi:10.1103/PhysRevD.90.045021
%[arXiv:1405.3789 [hep-th]].

\bibitem{cao}
G.~Cao, L.~He and P.~Zhuang,
%``Collective modes and Kosterlitz-Thouless transition in a magnetic field in the planar Nambu-Jona-Lasino model,''
Phys. Rev. D \textbf{90}, no.5, 056005 (2014).
%doi:10.1103/PhysRevD.90.056005
%[arXiv:1408.5364 [hep-ph]].

\bibitem{ekkz}
D.~Ebert, T.~G.~Khunjua, K.~G.~Klimenko and V.~C.~Zhukovsky,
%``Competition and duality correspondence between chiral and superconducting channels in ( 2+1 )-dimensional four-fermion models with fermion number and chiral chemical potentials,''
Phys. Rev. D \textbf{93}, no.10, 105022 (2016).
%doi:10.1103/PhysRevD.93.105022
%[arXiv:1603.00357 [hep-th]].

\bibitem{pauli}
W. Pauli, Nuovo Cimento, {\bf 6}, 204 (1957); F. Gursey, Nuovo Cimento, {\bf 7}, 411, (1957).

\bibitem{ojima}
  I.~Ojima and R.~Fukuda,
  %``Spontaneous Breakdown of Fermion Number Conservation in Two-Dimensions,''
  Prog.\ Theor.\ Phys.\  {\bf 57}, 1720 (1977).
  %%CITATION = PTPKA,57,1720;%%

\bibitem{kkz18}
T.~G.~Khunjua, K.~G.~Klimenko and R.~N.~Zhokhov,
%``Dense baryon matter with isospin and chiral imbalance in the framework of NJL$_4$ model at large $N_c$: duality between chiral symmetry breaking and charged pion condensation,''
Phys. Rev. D \textbf{97}, no.5, 054036 (2018).
%doi:10.1103/PhysRevD.98.054030
%[arXiv:1804.01014 [hep-ph]].
%doi:10.1103/PhysRevD.97.054036
%[arXiv:1710.09706 [hep-ph]].

\bibitem{kkz}
T.~G.~Khunjua, K.~G.~Klimenko and R.~N.~Zhokhov,
%``Charged pion condensation and duality in dense and hot chirally and isospin asymmetric quark matter in the framework of the NJL$_2$ model,''
Phys. Rev. D \textbf{100}, no.3, 034009 (2019).
%doi:10.1103/PhysRevD.100.034009
%[arXiv:1907.04151 [hep-ph]].

\bibitem{Thies2}
M.~Thies, %``Phase structure of the ( 1+1 )-dimensional Nambu\textendash{}Jona-Lasinio model with isospin,''
Phys. Rev. D \textbf{101}, no.1, 014010 (2020);
%``Duality study of the chiral Heisenberg-Gross-Neveu model in 1+1 dimensions,''
Phys. Rev. D \textbf{102}, no.9, 096006 (2020).
%doi:10.1103/PhysRevD.102.096006
%[arXiv:2008.13119 [hep-th]].

\bibitem{kkz18-2}
T.~G.~Khunjua, K.~G.~Klimenko and R.~N.~Zhokhov,
Phys. Rev. D \textbf{98}, no.5, 054030 (2018).
%doi:10.1103/PhysRevD.98.054030
%[arXiv:1804.01014 [hep-ph]].

\bibitem{kkz19}
T.~G.~Khunjua, K.~G.~Klimenko and R.~N.~Zhokhov,
%``Chiral imbalanced hot and dense quark matter: NJL analysis at the physical point and comparison with lattice QCD,''
Eur. Phys. J. C \textbf{79}, no.2, 151 (2019).
%doi:10.1140/epjc/s10052-019-6654-2
%[arXiv:1812.00772 [hep-ph]].

\bibitem{kkz20}
T.~G.~Khunjua, K.~G.~Klimenko and R.~N.~Zhokhov,
%``The dual properties of chiral and isospin asymmetric dense quark matter formed of two-color quarks,''
JHEP \textbf{06}, 148 (2020);
%doi:10.1007/JHEP06(2020)148
%[arXiv:2003.10562 [hep-ph]].
Phys. Part. Nucl. \textbf{53}, no.2, 461 (2022).

\bibitem{kogut}
  J.~B.~Kogut, M.~A.~Stephanov and D.~Toublan,
  %``On two color QCD with baryon chemical potential,''
  Phys.\ Lett.\ B {\bf 464}, 183 (1999);
  %doi:10.1016/S0370-2693(99)00971-5
 %[hep-ph/9906346];
  %%CITATION = doi:10.1016/S0370-2693(99)00971-5;%%
  J.~B.~Kogut, M.~A.~Stephanov, D.~Toublan, J.~J.~M.~Verbaarschot and A.~Zhitnitsky,
  %``QCD - like theories at finite baryon density,''
  Nucl.\ Phys.\ B {\bf 582}, 477 (2000).
  %doi:10.1016/S0550-3213(00)00242-X
  %[hep-ph/0001171].
  %%CITATION = doi:10.1016/S0550-3213(00)00242-X;%%

\bibitem{son2}
  K.~Splittorff, D.~T.~Son and M.~A.~Stephanov,
  %``QCD - like theories at finite baryon and isospin density,''
  Phys.\ Rev.\ D {\bf 64}, 016003 (2001).
  %doi:10.1103/PhysRevD.64.016003
 % [hep-ph/0012274].
  %%CITATION = doi:10.1103/PhysRevD.64.016003;%%

\bibitem{weise}
C.~Ratti and W.~Weise,
  %``Thermodynamics of two-colour QCD and the Nambu Jona-Lasinio model,''
  Phys.\ Rev.\ D {\bf 70}, 054013 (2004).
  %doi:10.1103/PhysRevD.70.054013
  %[hep-ph/0406159].
  %%CITATION = doi:10.1103/PhysRevD.70.054013;%%

\bibitem{ramos}
  D.~C.~Duarte, P.~G.~Allen, R.~L.~S.~Farias, P.~H.~A.~Manso, R.~O.~Ramos and N.~N.~Scoccola,
  %``BEC-BCS crossover in a cold and magnetized two color NJL model,''
  Phys.\ Rev.\ D {\bf 93}, 025017 (2016).
 
\bibitem{andersen3}
  J.~O.~Andersen and A.~A.~Cruz,
  Phys.\ Rev.\ D {\bf 88}, 025016 (2013).

  \bibitem{brauner1}
  T.~Brauner, K.~Fukushima and Y.~Hidaka,
Phys.\ Rev.\ D {\bf 80}, 074035 (2009)
  Erratum: [Phys.\ Rev.\ D {\bf 81}, 119904 (2010)].
 
\bibitem{andersen2}
  J.~O.~Andersen and T.~Brauner,
  Phys.\ Rev.\ D {\bf 81}, 096004 (2010).
 
\bibitem{imai}
S.~Imai, H.~Toki and W.~Weise,
Nucl.\ Phys.\ A {\bf 913}, 71 (2013).

\bibitem{chao}
  J.~Chao, Chin. Phys. C \textbf{44}, no.3, 034108 (2020).

\bibitem{Astrakhantsev:2020tdl}
N.~Astrakhantsev, V.~V.~Braguta, E.~M.~Ilgenfritz, A.~Y.~Kotov and A.~A.~Nikolaev,
%``Lattice study of thermodynamic properties of dense QC$_2$D,''
Phys. Rev. D \textbf{102}, no.7, 074507 (2020).

\bibitem{u2}
T.~G.~Khunjua, K.~G.~Klimenko and R.~N.~Zhokhov,
%``Dual properties of dense quark matter with color superconductivity phenomenon,''
Phys. Rev. D \textbf{108}, no.12, 125011 (2023).
%doi:10.1103/PhysRevD.108.125011
%[arXiv:2310.08211 [hep-ph]].

  \bibitem{andrianov}
 R.~Gatto and M.~Ruggieri,
Phys.\ Rev.\ D {\bf 85}, 054013 (2012);
  %%CITATION = ARXIV:1110.4904;%%
 L.~Yu, H.~Liu and M.~Huang,
Phys.\ Rev.\ D {\bf 90}, 074009 (2014);
Phys.\ Rev.\ D {\bf 94}, 014026 (2016);
  %%CITATION = ARXIV:1511.03073;%%
 M.~Ruggieri and G.~X.~Peng,
J.\ Phys.\ G {\bf 43}, no. 12, 125101 (2016);
A.~A.~Andrianov, V.~A.~Andrianov and D.~Espriu,
Particles {\bf 3}, no. 1, 15 (2020);
D.~Espriu, A.~G.~Nicola and A.~Vioque-Rodríguez, JHEP \textbf{06}, 062 (2020).
 % arXiv:2002.11696 [hep-ph].
  %%CITATION = ARXIV:2002.11696;%%
 
\bibitem{son}
D. T.~Son and M. A.~Stephanov, Phys.\ Atom.\ Nucl.\  {\bf 64}, 834 (2001);
D. C.~Duarte, R. L. S.~Farias and R. O.~Ramos,
  Phys.\ Rev.\  D {\bf 84}, 083525 (2011);
  %%CITATION = PHRVA,D84,083525;%%
D.~Ebert, K. G.~Klimenko, A. V.~Tyukov and V. C.~Zhukovsky,
  %``Pion condensation of quark matter in the static Einstein %universe,''
  Eur.\ Phys.\ J.\ C {\bf 58}, 57 (2008).
  %%CITATION = ARXIV:0804.0765;%%

\bibitem{he}
L. He, M. Jin, and P. Zhuang, Phys. Rev. D {\bf 71}, 116001 (2005);
 %\bibitem{eklim}
D. Ebert and K. G. Klimenko, J.\ Phys.\ G {\bf 32}, 599 (2006);
%%CITATION = JPHGB,G32,599;%%
Eur.\ Phys.\ J.\  C {\bf 46}, 771 (2006);
%%CITATION = EPHJA,C46,771;%%
C.f.~Mu, L.y.~He and Y.x.~Liu,
  Phys.\ Rev.\  D {\bf 82}, 056006 (2010).
  %%CITATION = PHRVA,D82,056006;%%

\bibitem{ak}
J. O.~Andersen and T.~Brauner,
  Phys.\ Rev.\  D {\bf 78}, 014030 (2008);
  %%CITATION = PHRVA,D78,014030;%%
J. O.~Andersen and L.~Kyllingstad,
 J.\ Phys.\ G {\bf 37}, 015003 (2009);
P.~Adhikari, J.~O.~Andersen and P.~Kneschke,
Phys.\ Rev.\ D {\bf 98},  074016  (2018);
Eur.\ Phys.\ J.\ C {\bf 79},  874 (2019);
J.~O.~Andersen, P.~Adhikari and P.~Kneschke,
PoS \textbf{Confinement2018}, 197 (2019)
%doi:10.22323/1.336.0197
[arXiv:1810.00419 [hep-ph]].
%arXiv:1810.00419 [hep-ph].

\bibitem{ekkz2}
 D.~Ebert, T. G.~Khunjua, K. G.~Klimenko and V. C.~Zhukovsky,
  %``Charged pion condensation phenomenon of dense baryonic matter induced by finite volume: The NJL(2) model consideration,''
  Int.\ J.\ Mod.\ Phys.\ A {\bf 27}, 1250162 (2012);
  %%CITATION = doi:10.1142/S0217751X1250162X;%%
  %\bibitem{gkkz}
  N. V.~Gubina, K. G.~Klimenko, S. G.~Kurbanov and V. C.~Zhukovsky,
  Phys.\ Rev.\ D {\bf 86}, 085011 (2012).
  %%CITATION = doi:10.1103/PhysRevD.86.085011;%%

\bibitem{Andersen:2018nzq}
 J.~O.~Andersen and P.~Kneschke,
 %``Bose-Einstein condensation and pion stars,''
arXiv:1807.08951 [hep-ph];
B.~B.~Brandt, G.~Endrodi, E.~S.~Fraga, M.~Hippert, J.~Schaffner-Bielich and S.~Schmalzbauer,
Phys.\ Rev.\ D {\bf 98},  094510 (2018).
       
\bibitem{Ayala}
A.~Ayala, A.~Bandyopadhyay, R.~L.~S.~Farias, L.~A.~Hern\'andez and J.~L.~Hern\'andez,
Phys. Rev. D \textbf{107}, no.7, 074027 (2023).
%arXiv:2301.13633 [hep-ph].       

\bibitem{fukus}
K. Fukushima, D. E. Kharzeev and H. J. Warringa, Phys.Rev.D {\bf 78}, 074033 (2008).

\bibitem{Metlitski} 
M.~A.~Metlitski and A.~R.~Zhitnitsky,
Phys.\ Rev.\ D {\bf 72}, 045011 (2005).

\bibitem{Khun}
T.~G.~Khunjua, K.~G.~Klimenko and R.~N.~Zhokhov,
JHEP \textbf{06}, 006 (2019).

\bibitem{iliopoulos}
J.~Iliopoulos, C.~Itzykson and A.~Martin,
%``Functional Methods and Perturbation Theory,''
Rev. Mod. Phys. \textbf{47}, 165 (1975).

\bibitem{u}
T.~G.~Khunjua, K.~G.~Klimenko and R.~N.~Zhokhov, 
Phys. Rev. D \textbf{106}, no.4, 045008 (2022).


\bibitem{buballa}
M. Buballa, Phys. Rep. {\bf 407}, 205 (2005);
%\bibitem{alford}
I. A. Shovkovy, Found. Phys. {\bf 35}, 1309 (2005);
 M.~Huang,
  %``Color superconductivity at moderate baryon density,''
  Int.\ J.\ Mod.\ Phys.\ E {\bf 14}, 675 (2005);
 % doi:10.1142/S0218301305003491
 % [hep-ph/0409167].
  %%CITATION = doi:10.1142/S0218301305003491;%%
 K.~G.~Klimenko and D.~Ebert,
  %``Mesons and diquarks in a dense quark medium with color superconductivity,''
  Theor.\ Math.\ Phys.\  {\bf 150}, 82 (2007) [Teor.\ Mat.\ Fiz.\  {\bf 150}, 95 (2007)];
  %doi:10.1007/s11232-007-0006-3
  %%CITATION = doi:10.1007/s11232-007-0006-3;%%
 M. G.~Alford, A.~Schmitt, K.~Rajagopal, and T.~Sch\"afer,
 Rev.\ Mod.\ Phys.\  {\bf 80}, 1455 (2008);
  %%CITATION = RMPHA,80,1455;%%
 E.~J.~Ferrer and V.~de la Incera,
  %``Magnetism in Dense Quark Matter,''
  Lect.\ Notes Phys.\  {\bf 871}, 399 (2013).
 % doi:10.1007/978-3-642-37305-3_16
 % [arXiv:1208.5179 [nucl-th]].
  %%CITATION = doi:10.1007/978-3-642-37305-3_16;%%  

%\cite{Smilga:1994tb}
\bibitem{Smilga:1994tb}
A.~V.~Smilga and J.~J.~M.~Verbaarschot,
%``Spectral sum rules and finite volume partition function in gauge theories with real and pseudoreal fermions,''
Phys. Rev. D \textbf{51}, 829 (1995).
%doi:10.1103/PhysRevD.51.829
%[arXiv:hep-th/9404031 [hep-th]].
%114 citations counted in INSPIRE as of 25 Nov 2024



\end{thebibliography}
\end{document}